\documentclass[a4paper,graphics,natbib,graphicx,psfig,amssymb,ccaption]{article}
\usepackage{lscape}
\usepackage{graphicx}
\newcommand{\affil}[1]{$^{\rm #1}$}
\usepackage[authoryear]{natbib}
\bibpunct{(}{)}{;}{a}{}{,}
\date{} 

\title{\large\bf\flushleft Absolute Magnitude Calibration for Red Giants based on the Colour -- Magnitude Diagrams of Galactic Clusters. I- Calibration with {\em V} and {\em B-V}}
\author{\parbox{\textwidth}{\flushleft
\vspace{-0.5cm}
{\it S. Karaali\affil{\dag, A, B}, S. Bilir\affil{A}, and E. Yaz G\"ok\c ce\affil{A}}\\
\vspace{0.4cm}
{\small \affil{A}\,Istanbul University, Faculty of Sciences, Department of Astronomy and Space Sciences, 34119, Istanbul, Turkey}\\
{\small \affil{B}\,Email: karsa@istanbul.edu.tr}}}
\begin{document}
\twocolumn[
\begin{changemargin}{.8cm}{.5cm}
\begin{minipage}{.9\textwidth}
\vspace{-1cm}
\maketitle
\small{\bf Abstract:}
We present an absolute magnitude calibration for red giants with the colour magnitude diagrams of six Galactic clusters with different metallicities i.e. M92, M13, M5, 47 Tuc, M67, and NGC 6791. The combination of the absolute magnitude offset from the fiducial of giant sequence of the cluster M5 with the corresponding metallicity offset provides calibration for absolute magnitude estimation for red giants for a given $(B-V)_{0}$ colour. The calibration is defined in the colour interval $0.75\leq(B-V)_{0}\leq1.50$ mag and it covers the metallicity interval $-2.15<\lbrack Fe/H\rbrack \leq+0.37$ dex. $91\%$ of the absolute magnitude residuals obtained by the application of the procedure to another set of Galactic clusters lie in the interval $-0.40<\Delta M\leq+0.40$ mag. The mean and the standard deviation of the residuals are 0.05 and 0.19 mag, respectively. We fitted the absolute magnitude also to metallicity and age for a limited sub-sample of $(B-V)_{0}$ colour, just to test the effect of age in absolute magnitude calibration. Comparison of the mean and the standard deviation of the residuals evaluated by this procedure with the corresponding ones provided by the procedure where the absolute magnitude fitted to a third degree polynomial of metallicity show that the age parameter may be omitted in absolute magnitude estimation of red giants. The derived relations are applicable to stars older than 4 Gyr, the age of the youngest calibrating cluster.    

\medskip
\medskip
\end{minipage}
\end{changemargin}
]
\small
\let\thefootnote\relax\footnote{\small \affil{\dag}\,Retired.}
\section{Introduction}
Stellar kinematics and metallicity are two primary means to deduce the history of our Galaxy. However, such goals can not be achieved without stellar distances. The distance to a star can be evaluated by trigonometric or photometric parallaxes. Trigonometric parallaxes are only available for nearby stars where {\em Hipparcos} \citep{Perryman97} is the main supplier for the data. For stars at large distances, the use of photometric parallaxes is unavoidable. In other words the study of the Galactic structure is strictly tied to precise  determination of absolute magnitudes.

Different methods can be used for absolute magnitude determination where most of them  are devoted to dwarfs. The method used in the Str\"omgren's $uvby-\beta$ \citep{NS91} and in the {\em UBV} \citep{Laird88} photometries depends on the absolute magnitude offset from a standard main-sequence. In recent years the derivation of absolute magnitudes has been carried out by means of colour-absolute magnitude diagrams of some specific clusters whose metal abundances are generally adopted as the mean metal abundance for a Galactic population, such as thin, thick discs and halo. The studies of \cite{Phleps00} and \cite{Chen01} can be given as examples. A slightly different approach is that of \citet{Siegel02} where two relations, one for stars with solar-like abundances and another one for metal-poor stars were derived between $M_{R}$ and the colour index $R-I$, where $M_{R}$ is the absolute magnitude in the R filter of Johnson system. For a star of given metallicity and colour, absolute magnitude can be estimated by {\em linear} interpolation of {\em two} ridgelines and by means of {\em linear} extrapolation beyond the metal-poor ridgeline.

The most recent procedure used for absolute magnitude determination consists of finding the most likely values of the stellar parameters, given the measured atmospheric ones, and the time spent by a star in each region of the H-R diagram. In practice, researchers select the subset of isochrones with $[M/H]\pm \Delta_{[M/H]}$, where $\Delta_{[M/H]}$ is the estimated error on the metallicity, for each set of derived $T_{eff}$, $\log g$ and $[M/H]$. Then a Gaussian weight is associated to each point of the selected isochrones, which depends on the measured atmospheric parameters and the considered errors. This criterion allows the algorithm to select only the points whose values are closed by the pipeline. For details of this procedure we cite the works of \cite{Breddels10} and \cite{Zwitter10}. This procedure is based on many parameters. Hence it provides absolute magnitudes with high accuracy. Also it can be applied to both dwarf and giant stars simultaneously.

In \cite{Karaali03}, we presented a procedure for the photometric parallax estimation of dwarf stars which depends on the absolute magnitude offset from the main-sequence of the Hyades cluster. In this study, we will use a similar procedure for the absolute magnitude estimation of red giants by using the apparent magnitude-colour diagrams of Galactic clusters with different metallicities. In Section 2 we present the data. The procedure used for calibration is given in Section 3, and Section 4 is devoted to summary and discussion.

\section{Data}
Six clusters with different metallicities, i.e. M92, M13, M5, 47 Tuc, M67, and NGC 6791, 
were selected for our program. The range of the metallicity given in iron abundance is 
$-2.15 \leq [Fe/H] \leq +0.37$ dex. The $V-M_{V}$ apparent distance modulus, 
$(V-M_{V})_{0}$ true distance modulus, $E(B-V)$ colour excess, and $[Fe/H]$ iron abundance 
are given in Table 1, whereas the $V$ and $B-V$ data are presented in Table 2. We adopted  
$R = A_{V}/E(B-V)=$ 3.1 to convert between colour excess and extinction. Although different 
numerical values appeared in the literature for specific regions of our Galaxy, a single value 
is applicable everywhere. Different distance moduli and interstellar extinctions were cited in 
the literature for the clusters. The data in Table 1 and Table 2 are taken from the authors cited 
in the reference list of Table 1. The $V$ and $B-V$ data of M92 were taken from two sources, i.e.
 \cite{Sandage70} and \cite{Stetson88}, and combined to obtain a colour magnitude diagram with 
largest range in $B-V$. Data used in the combination of the final colour-magnitude diagram are 
shown with bold face in Table 2. The distance modulus, reddening, and metallicity for this cluster 
were taken from \cite{Gratton97}. All the data  for each cluster M5, M67 and NGC6791 were taken 
from a single source, as indicated in Table 1. Whereas, we refer two references for the clusters 
M13 and 47 Tuc. The $V$ and $B-V$ data were taken from the first reference, but the second one 
refers to their $V-M_{V}$ distance modulus, $E(B-V)$ colour-excess and $[Fe/H]$ metallicity. 
The reason of this selection is to obtain the best fitting of the colour magnitude diagrams to 
the isochrones (see section 3). Thus, in the case of more than one reference in Table1, the last 
one refers to the distance modulus, colour excess, and metallicity. The original $V$ and $B-V$ 
data refer to the fiducial sequence, i.e. giants, sub-giants, and main-sequence stars of the clusters. 
We plotted these sequences on a diagram for each cluster and identified red giants by means of their 
positions in the diagram. The $(V, B-V)$ points in Table 2 consist of the fiducial sequence of the 
referred cluster. Hence, they represent  the cluster in question quite well. However, they are not error free. 
The errors for these couples may be a bit larger for the photographic magnitude and colours of the clusters 
M92 and M13 than the CCD ones of the other clusters. As noted above the bright magnitudes and the corresponding 
colours of the cluster M92, and all magnitude and colours of the cluster M13 which were taken from \cite{Sandage70} 
are photographic data. We will see in Section 3 that the data of these clusters are in good 
agreement with the data of the other clusters investigated by using CCD technic. We, then fitted 
the fiducial sequence of the red giants to a sixth degree polynomial for all clusters, except M92 for which a seventh degree polynomial  was necessary for a good correlation coefficient. The calibration of $V_{0}$ is as follows:

\begin{table}
\setlength{\tabcolsep}{2pt}
\center
 \caption{Data for the clusters used in our work.}
\scriptsize{
 \label{tabledata}
 \begin{tabular}{lcccccl}
\hline
    Cluster  & $V-M_{V}$ & $E(B-V)$ & $(V-M_{V})_{0}$ & $[Fe/H]$ & $[\alpha/Fe]$ &Reference \\
             & (mag) & (mag) & (mag) & (dex) & & \\
\hline
    M92    & 14.80 & 0.025 & 14.72 & -2.15 &0.33 & (1),(2),(3)\\
    M13    & 14.44 & 0.020 & 14.38 & -1.41 &0.22 & (1),(3) \\
    M5     & 14.50 & 0.020 & 14.44 & -1.17 &0.24 & (4) \\
    47 Tuc & 13.37 & 0.040 & 13.25 & -0.80 &0.27 & (5),(6) \\
    M67    &  9.65 & 0.038 &  9.53 & -0.04 & $-$ & (7) \\
    NGC6791& 13.25 & 0.100 & 12.94 &  0.37 & $-$ & (7) \\
\hline
\end{tabular}\\
(1) \cite{Sandage70}, (2) \cite{Stetson88}, (3)\cite{Gratton97}, (4) \cite{Sandquist96}, (5) \cite{Hesser87}, (6) \cite{Percival02}, (7) \cite{Sandage03}
\label{tab:addlabel}
}
\end{table}

\begin{table*}
  \center
\setlength{\tabcolsep}{1.5pt}
  \caption{Fiducial red giant sequences for six Galactic clusters used for calibration. The data of M92 are taken from two different sources and combined to obtain a diagram with largest $(B-V)$-interval. The colours and magnitudes used for the calibration are given with bold face. 10 and 12 fiducial points of M67 and NGC 6791, respectively, in the original catalogues which correspond to sub-giants/main-sequence stars have not been considered in our work.}
\scriptsize{
    \begin{tabular}{cccc|cccc|cccc|cccc}
    \hline
    $B-V$ & $V$ & $(B-V)_{0}$ & $V_{0}$ & $B-V$ & $V$ & $(B-V)_{0}$ & $V_{0}$ & $B-V$ & $V$ & $(B-V)_{0}$ & $V_{0}$ & $B-V$ & $V$ & $(B-V)_{0}$ & $V_{0}$ \\
    \hline
\multicolumn{4}{c|}{M92 \citep{Stetson88}} & \multicolumn{4}{c|}{M92 \citep{Sandage70}} & \multicolumn{4}{c|}{M5 (cont.)} & \multicolumn{4}{c}{M67 (cont.)} \\
    \hline
    0.690 & 15.80 & 0.665 & 15.72 & \textbf{1.400} & \textbf{11.90} & \textbf{1.375} & \textbf{11.82} & 1.023 & 13.91 & 1.003 & 13.85 & 0.937 & 12.86 & 0.899 & 12.74 \\
    0.677 & 16.00 & 0.652 & 15.92 & \textbf{1.300} & \textbf{12.09} & \textbf{1.275} & \textbf{12.01} & 1.003 & 14.09 & 0.983 & 14.03 & 0.900 & 13.05 & 0.862 & 12.93 \\
    0.664 & 16.20 & 0.639 & 16.12 & \textbf{1.200} & \textbf{12.39} & \textbf{1.175} & \textbf{12.31} & 0.965 & 14.30 & 0.945 & 14.24 & 0.850 & 13.05 &  $-$    & $-$ \\
    0.652 & 16.40 & 0.627 & 16.32 & \textbf{1.100} & \textbf{12.75} & \textbf{1.075} & \textbf{12.67} & 0.903 & 14.70 & 0.883 & 14.64 & 0.800 & 12.90 &   $-$    & $-$ \\
    \textbf{0.639} & \textbf{16.60} & \textbf{0.614} & \textbf{16.52} & \textbf{1.000} & \textbf{13.15} & \textbf{0.975} & \textbf{13.07} & 0.867 & 14.94 & 0.847 & 14.87 & 0.750 & 12.75 &   $-$    & $-$ \\
    \textbf{0.627} & \textbf{16.80} & \textbf{0.602} & \textbf{16.72} & \textbf{0.900} & \textbf{13.58} & \textbf{0.875} & \textbf{13.50} & 0.855 & 15.09 & 0.835 & 15.03 & 0.700 & 12.63 &   $-$    &  $-$\\
    \textbf{0.615} & \textbf{17.00} & \textbf{0.590} & \textbf{16.92} & \textbf{0.800} & \textbf{14.11} & \textbf{0.775} & \textbf{14.03} & 0.831 & 15.32 & 0.811 & 15.26 & 0.650 & 12.61 &   $-$    &  $-$\\
    \textbf{0.603} & \textbf{17.20} & \textbf{0.578} & \textbf{17.12} & 0.750 & 14.45 & 0.725 & 14.37 & 0.814 & 15.50 & 0.794 & 15.44 & 0.600 & 12.60 &  $-$     &  $-$ \\
    \textbf{0.591} & \textbf{17.40} & \textbf{0.566} & \textbf{17.32} & 0.700 & 15.05 & 0.675 & 14.97 & 0.798 & 15.70 & 0.778 & 15.63 & 0.535 & 12.85 &  $-$     &  $-$\\
\cline{5-8}    \textbf{0.578} & \textbf{17.60} & \textbf{0.553} & \textbf{17.52} & \multicolumn{4}{c|}{M92 (Combined Sequence)} & 0.783 & 15.89 & 0.763 & 15.83 & 0.578 & 13.15 &  $-$     &  $-$\\
\cline{5-8}0.552 & 17.80 & 0.527 & 17.72 &$-$&$-$&1.375 & 11.82 & 0.773 & 16.10 & 0.753 & 16.04 & 0.559 & 13.50 &  $-$  & $-$ \\
    0.500 & 17.96 & 0.475 & 17.88 & $-$& $-$ & 1.275 & 12.01 & 0.756 & 16.30 & 0.736 & 16.24 & 0.562 & 13.75 &  $-$     & $-$\\
\cline{13-16}0.476 & 18.00 & 0.451 & 17.92 &$-$ &$-$&1.175&12.31&0.746 & 16.48 & 0.726 & 16.42 & \multicolumn{4}{c}{NGC 6791}  \\
\cline{13-16}0.450 & 18.08 & 0.425 & 18.00 &$-$ &$-$&1.075&12.67&0.731 & 16.70 & 0.711 & 16.63 & 1.535 & 14.25 & 1.435 & 13.94 \\
    0.419 & 18.20 & 0.394 & 18.12 &$-$ &$-$& 0.975 & 13.07 & 0.724 & 16.92 & 0.704 & 16.86 & 1.500 & 14.45 & 1.400 & 14.14 \\
    0.400 & 18.32 & 0.375 & 18.24 &$-$ &$-$& 0.875 & 13.50 & 0.711 & 17.10 & 0.691 & 17.04 & 1.450 & 14.73 & 1.350 & 14.42 \\
    0.396 & 18.40 & 0.371 & 18.32 &$-$ &$-$& 0.775 & 14.03 & 0.714 & 17.10 & 0.694 & 17.04 & 1.400 & 15.04 & 1.300 & 14.73 \\
    0.388 & 18.60 & 0.363 & 18.52 &$-$ &$-$& 0.614 & 16.52 & 0.717 & 17.30 & 0.697 & 17.24 & 1.331 & 15.50 & 1.231 & 15.19 \\
    0.390 & 18.80 & 0.365 & 18.72 &$-$ &$-$& 0.602 & 16.72 & 0.694 & 17.50 & 0.674 & 17.44 & 1.274 & 16.00 & 1.174 & 15.69 \\
    0.397 & 19.00 & 0.372 & 18.92 &$-$ &$-$& 0.590 & 16.92 & 0.684 & 17.65 & 0.664 & 17.59 & 1.228 & 16.50 & 1.128 & 16.19 \\
\cline{9-12}0.406 & 19.20 & 0.381 & 19.12 &$-$ &$-$& 0.578 & 17.12 & \multicolumn{4}{c|}{47 Tuc}& 1.191 & 17.00 & 1.091 & 16.69 \\
\cline{9-12}0.418 & 19.40 & 0.393 & 19.32 &$-$ &$-$& 0.566 & 17.32 & 1.70& 11.70 & 1.66 & 11.58 & 1.175 & 17.25 & 1.075 & 16.94 \\
    0.434 & 19.60 & 0.409 & 19.52 &$-$ &$-$& 0.553 & 17.52 & 1.60  & 11.85 & 1.56  & 11.73 & 1.140 & 17.50 & 1.040 & 17.19 \\
\cline{5-8}0.454 & 19.80 & 0.429 & 19.72 & \multicolumn{4}{c|}{M13} & 1.50  & 12.03 & 1.46  & 11.91 & 1.100 & 17.50 & $-$ & $-$ \\
\cline{5-8}0.477 & 20.00 & 0.452 & 19.92 & 1.62 & 12.05 & 1.60 & 11.99 & 1.40 & 12.23 & 1.36 & 12.11 & 1.050 &17.46 & $-$ & $-$ \\
    0.502 & 20.20 & 0.477 & 20.12 & 1.50  & 12.05 & 1.48  & 11.99 & 1.30  & 12.55 & 1.26  & 12.43 & 1.000 & 17.39 & $-$ & $-$ \\
    0.529 & 20.40 & 0.504 & 20.32 & 1.40  & 12.16 & 1.38  & 12.10 & 1.19  & 13.00 & 1.15  & 12.88 & 0.950 & 17.35 & $-$ & $-$ \\
    0.559 & 20.60 & 0.534 & 20.52 & 1.30  & 12.40 & 1.28  & 12.34 & 1.10  & 13.50 & 1.06  & 13.38 & 0.900 & 17.42 & $-$ & $-$ \\
    0.594 & 20.80 & 0.569 & 20.72 & 1.20  & 12.70 & 1.18  & 12.64 & 1.01  & 14.00 & 0.97  & 13.88 & 0.888 & 17.50 & $-$ & $-$ \\
    0.632 & 21.00 & 0.607 & 20.92 & 1.10  & 13.09 & 1.08  & 13.03 & 0.95  & 14.50 & 0.91  & 14.38 & 0.875 & 17.75 & $-$ & $-$ \\
    0.675 & 21.20 & 0.650 & 21.12 & 1.00  & 13.55 & 0.98  & 13.49 & 0.90  & 15.00 & 0.86  & 14.88 & 0.888 & 18.00 & $-$ & $-$ \\
    0.718 & 21.40 & 0.693 & 21.32 & 0.95  & 13.85 & 0.93  & 13.79 & 0.83  & 16.00 & 0.79  & 15.88 & 0.909 & 18.25 & $-$ & $-$ \\
    0.763 & 21.60 & 0.738 & 21.52 & 0.90  & 14.15 & 0.88  & 14.09 & 0.81  & 16.50 & 0.77  & 16.38 & 0.942 & 18.50 & $-$ & $-$ \\
    0.810 & 21.80 & 0.785 & 21.72 & 0.85  & 14.64 & 0.83  & 14.58 & 0.79  & 17.00 & 0.75  & 16.88 & 0.974 & 18.75 & $-$ & $-$ \\
\cline{5-12}0.855 & 22.00 & 0.830 & 21.92 & \multicolumn{4}{c|}{M5 } & \multicolumn{4}{c|}{M67}& 1.010 & 19.00 & $-$ & $-$ \\
\cline{5-12}0.901 & 22.20 & 0.876 & 22.12 & 1.653 & 12.06 & 1.633 & 11.99 & 1.555 & 9.00  & 1.517 & 8.88  & $-$ & $-$ & $-$ & $-$ \\
    0.949 & 22.40 & 0.924 & 22.32 & 1.582 & 12.17 & 1.562 & 12.11 & 1.430 & 9.50  & 1.392 & 9.38  &  $-$ & $-$ & $-$ & $-$ \\
    0.996 & 22.60 & 0.971 & 22.52 & 1.483 & 12.30 & 1.463 & 12.24 & 1.320 & 10.00 & 1.282 & 9.88  &  $-$ & $-$ & $-$ & $-$ \\
    1.042 & 22.80 & 1.017 & 22.72 & 1.416 & 12.46 & 1.396 & 12.39 & 1.220 & 10.50 & 1.182 & 10.38 &  $-$ & $-$ & $-$ & $-$ \\
    1.074 & 23.00 & 1.049 & 22.92 & 1.342 & 12.63 & 1.322 & 12.57 & 1.127 & 11.00 & 1.089 & 10.88 &  $-$ & $-$ & $-$ & $-$ \\
    1.089 & 23.20 & 1.064 & 23.12 & 1.253 & 12.90 & 1.233 & 12.84 & 1.059 & 11.50 & 1.021 & 11.38 &  $-$ & $-$ & $-$ & $-$ \\
    1.092 & 23.40 & 1.067 & 23.32 & 1.194 & 13.10 & 1.174 & 13.04 & 1.032 & 11.75 & 0.994 & 11.63 &  $-$ & $-$ & $-$ & $-$ \\
    1.086 & 23.60 & 1.061 & 23.52 & 1.141 & 13.32 & 1.121 & 13.26 & 1.000 & 12.18 & 0.962 & 12.06 &  $-$ & $-$ & $-$ & $-$ \\
    1.077 & 23.80 & 1.052 & 23.72 & 1.092 & 13.52 & 1.072 & 13.46 & 0.974 & 12.50 & 0.936 & 12.38 &  $-$ & $-$ & $-$ & $-$ \\
    1.060 & 24.00 & 1.035 & 23.92 & 1.053 & 13.72 & 1.033 & 13.66 & 0.950 & 12.72 & 0.912 & 12.60 &  $-$ & $-$ & $-$ & $-$ \\
       \hline
    \end{tabular}
}
  \label{cl-calibration}
\end{table*}

\begin{figure}
\begin{center}
\includegraphics[scale=0.50, angle=0]{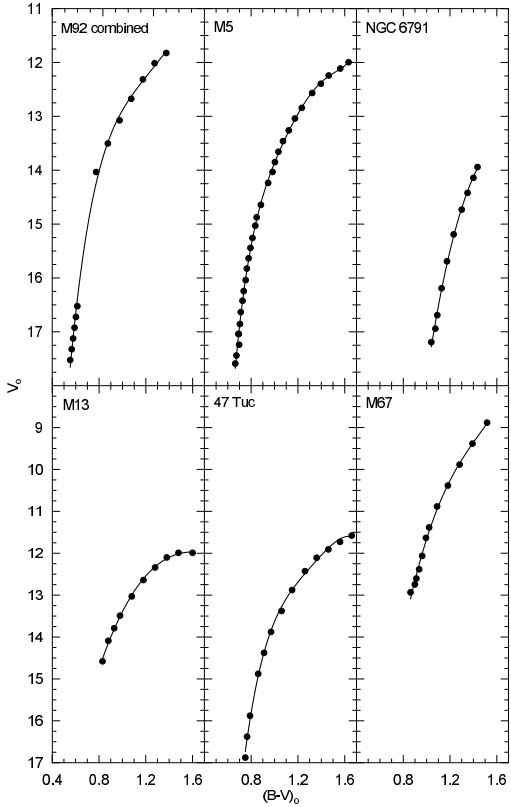} 
\caption[] {$V_{0}\times(B-V)_{0}$ colour-apparent magnitude diagrams for six Galactic clusters used for the absolute magnitude calibration.} 
\label{his:col}
\end{center}
\end {figure}

\begin{table}[h]
\setlength{\tabcolsep}{2.5pt}
  \center
\tiny{
  \caption{Numerical values of the coefficients $a_{i}$ ($i=0,1,2,3,4, 5, 6, 7$) in Eq. (1).}
    \begin{tabular}{lrrrrrr}
    \hline
    Cluster & M92   & M13   & M5    & 47 Tuc & M67   & NGC 6791 \\
    \hline
$(B-V)_{0}$  & [0.55, 1.38] & [0.83, 1.60] & [0.66, 1.63] & [0.75, 1.66] & [0.86, 1.52] & [1.04, 1.44] \\
interval &  &  &  &  &  & \\
    $a_{7}$ &   1260.209 & $-$ & $-$ & $-$ & $-$ & $-$ \\
    $a_{6}$ &  -8985.514 &   98.924 &    6.240 &   143.690 &  -1353.155 & - 12852.318 \\
    $a_{5}$ &  27088.160 & -753.362 &  -76.946 & -1101.784 &   9923.884 &   97469.551 \\
    $a_{4}$ & -44677.194 & 2376.106 &  332.511 &  3484.807 & -30101.159 & -307314.819 \\
    $a_{3}$ &  43443.029 &-3970.848 & -700.971 & -5820.763 &  48311.495 &  515575.766 \\
    $a_{2}$ & -24830.833 & 3710.911 &  789.927 &  5421.810 & -43245.725 & -485363.694 \\
    $a_{1}$ &   7688.065 &-1845.252 & -465.120 & -2679.997 &  20450.384 &  243057.102 \\
    $a_{0}$ &   -972.939 &  396.915 &  128.221 &   565.960 &  -3974.116 &  -50554.897 \\
    \hline
    \end{tabular}
}
  \label{tab:addlabel}
\end{table}

\begin{eqnarray}
   V_{0}= \sum_{i=1}^{7}a_{i}(B-V)^{i}_0 
\end{eqnarray}
The numerical values of the coefficients $a_{i}$ ($i =$ 0, 1, 2, 3, 4, 5, 6, 7) are given in Table 3, and the corresponding diagrams are presented in Fig. 1. The $(B-V)_{0}$-interval in the first line of the table indicate to the range of $(B-V)_{0}$ available for each cluster.

\begin{figure}
\begin{center}
\includegraphics[scale=0.60, angle=0]{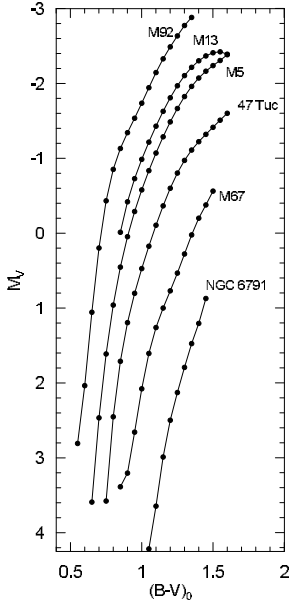} 
\caption[] {$M_{V}\times(B-V)_{0}$ colour-absolute magnitude diagrams for six Galactic clusters used for the absolute magnitude calibration.} 
\label{his:col}
\end{center}
\end {figure}

\section{The Procedure}
\subsection{Absolute Magnitude Offset as a Function of Metallicity Offset}
The procedure consists of calibration of the absolute magnitude offsets from the fiducial giant sequence of a standard cluster as a function of metallicity offsets. For this purpose we proceeded as in the following. We estimated the absolute magnitudes for the $(B-V)_{0}$ colours given in Table 4 for the cluster sample by combining the $V_{0}$ apparent magnitudes evaluated by Eq. (1) and the true distance modulus ($\mu_{0}$) of the cluster in question, i.e.
\begin{eqnarray}
M_{V}=V_{0}-\mu_{0}.     
\end{eqnarray}                                  
Then, we plotted the absolute magnitudes versus $(B-V)_{0}$ colours. Fig. 2 shows that the absolute magnitude is colour and metallicity dependent. It increases (algebraically) with increasing metallicity and decreasing colour. We fitted the $M_{V}-(B-V)_{0}$ diagrams to isochrones in order to test our data and the procedure. The diagrams of five clusters with metallicity  $[Fe/H] \leq$ 0.2 dex could be fitted to the Padova isochrones \citep{Marigo08}. Whereas it could not be carried out for the cluster NGC 6791 whose metallicity is beyond the upper metallicity limit of the Padova isochrones, i.e. $[Fe/H]=0.2$ dex. Hence, we used the isochrones of the Victoria-Regina Stellar Models \citep{VandenBerg06} for the cluster NGC 6791. The results are given in Fig. 3. There is a good fitting of the fiducial sequence of the red giants to the isochrones for a large interval of the $(B-V)_{0}$ colour index. However, one can notice a deviation for the metal-poor clusters at the red segment. We should note that the fitting presented in Fig. 3, for each cluster, is the best one of different combinations of distance modulus, reddening, and metallicity. We adopted the sequence of M5 as the standard one and we evaluated the $\Delta M$ offsets from the fiducial giant sequence of M5 (Table 4). Now, we can replace $\Delta M$ versus the corresponding $\Delta[Fe/H]$ iron abundance residuals and obtain the required calibration. This is carried out for the colours $(B-V)_{0}=$ 0.75, 1.00, 1.25, and 1.50 mag just to test of the procedure, and the results are given in Table 5 and Fig. 4. This procedure can be applied to any $(B-V)_{0}$ colour-interval for which the sample clusters are defined. 

\begin{table*}
\setlength{\tabcolsep}{5pt}
  \center
  \caption{$M_{V}$ absolute magnitudes and $\Delta M$ offsets estimated for a set of $(B-V)_{0}$ interval for six Galactic clusters used in the calibration.}
    \begin{tabular}{ccccccccccccc}
    \hline
 & \multicolumn{2}{c}{M92} & \multicolumn{2}{c}{M13} & \multicolumn{2}{c}{M5} & \multicolumn{2}{c}{47 Tuc} & \multicolumn{2}{c}{M67} & \multicolumn{2}{c}{NGC 6791} \\
    \hline
$(B-V)_{0}$ & $M_{V}$ &  $\Delta M$ & $M_{V}$ &  $\Delta M$ & $M_{V}$ &  $\Delta M$ & $M_{V}$ & $\Delta M$ & $M_{V}$ & $\Delta M $ & $M_{V}$ &  $\Delta M$\\  
\hline
    0.55  & 2.81  & $-$   &  $-$  &  $-$  &  $-$  & $-$   & $-$   & $-$   & $-$   & $-$   & $-$   & $-$ \\
    0.60  & 2.04  & $-$   &  $-$  &  $-$  &  $-$  & $-$   & $-$   & $-$   & $-$   & $-$   & $-$   & $-$ \\
    0.65  & 1.06  & -2.54 &  $-$  &  $-$  & 3.59  & 0     & $-$   & $-$   & $-$   & $-$   & $-$   & $-$ \\
    0.70  & 0.20  & -2.27 &  $-$  &  $-$  & 2.47  & 0     & $-$   & $-$   & $-$   & $-$   & $-$   & $-$ \\
    0.75  & -0.43 & -2.05 &  $-$  &  $-$  & 1.61  & 0     & 3.58  & 1.96  & $-$   & $-$   & $-$   & $-$ \\
    0.80  & -0.85 & -1.81 &  $-$  &  $-$  & 0.96  & 0     & 2.45  & 1.49  & $-$   & $-$   & $-$   & $-$ \\
    0.85  & -1.13 & -1.58 & -0.01 & -0.47 & 0.45  & 0     & 1.71  & 1.26  & 3.39  & 2.93  & $-$   & $-$ \\
    0.90  & -1.34 & -1.39 & -0.42 & -0.47 & 0.05  & 0     & 1.19  & 1.15  & 3.20  & 3.16  & $-$   & $-$ \\
    0.95  & -1.54 & -1.25 & -0.73 & -0.44 & -0.29 & 0     & 0.80  & 1.09  & 2.66  & 2.95  & $-$   & $-$ \\
    1.00  & -1.74 & -1.16 & -0.99 & -0.41 & -0.58 & 0     & 0.47  & 1.05  & 2.08  & 2.66  & $-$   & $-$ \\
    1.05  & -1.94 & -1.11 & -1.22 & -0.38 & -0.84 & 0     & 0.17  & 1.01  & 1.61  & 2.44  & 4.22  & 5.05 \\
    1.10  & -2.15 & -1.08 & -1.43 & -0.36 & -1.07 & 0     & -0.11 & 0.96  & 1.26  & 2.33  & 3.64  & 4.72 \\
    1.15  & -2.33 & -1.04 & -1.63 & -0.34 & -1.29 & 0     & -0.37 & 0.92  & 1.00  & 2.29  & 2.99  & 4.28 \\
    1.20  & -2.49 & -1.00 & -1.81 & -0.32 & -1.49 & 0     & -0.60 & 0.89  & 0.77  & 2.26  & 2.50  & 3.98 \\
    1.25  & -2.64 & -0.97 & -1.97 & -0.30 & -1.67 & 0     & -0.80 & 0.86  & 0.53  & 2.20  & 2.13  & 3.79 \\
    1.30  & -2.77 & -0.95 & -2.11 & -0.28 & -1.83 & 0     & -0.97 & 0.85  & 0.28  & 2.10  & 1.79  & 3.62 \\
    1.35  & -2.88 & -0.92 & -2.22 & -0.26 & -1.96 & 0     & -1.11 & 0.85  & 0.02  & 1.98  & 1.47  & 3.44 \\
    1.40  &  $-$  &  $-$  & -2.30 & -0.23 & -2.07 & 0     & -1.22 & 0.85  & -0.20 & 1.87  & 1.20  & 3.28 \\
    1.45  &  $-$  &  $-$  & -2.37 & -0.20 & -2.16 & 0     & -1.32 & 0.84  & -0.38 & 1.79  & 0.87  & 3.04 \\
    1.50  &  $-$  &  $-$  & -2.41 & -0.17 & -2.24 & 0     & -1.42 & 0.82  & -0.56 & 1.68  & $-$   & $-$ \\
    1.55  &  $-$  &  $-$  & -2.42 & -0.12 & -2.31 & 0     & -1.51 & 0.80  &  $-$  & $-$   & $-$   & $-$ \\
    1.60  &  $-$  &  $-$  & -2.39 & -0.01 & -2.38 & 0     & -1.60 & 0.78  &  $-$  & $-$   & $-$   & $-$ \\

\hline
    \end{tabular}
  \label{tab:addlabel}
\end{table*}

\begin{table}
  \center
  \caption{$\Delta M$ and $\Delta[Fe/H]$ offsets for four $(B-V)_{0}$ intervals.}
    \begin{tabular}{ccc}
    \hline
    $(B-V)_{0}$ & \multicolumn{1}{c}{ $\Delta[Fe/H]$} & \multicolumn{1}{c}{$\Delta M$} \\
    \hline
    0.75  & -0.98 & -2.05 \\
          &  0.00 &  0.00 \\
          &  0.37 &  1.96 \\
    \hline
    1.00  & -0.98 & -1.16 \\
          & -0.24 & -0.41 \\
          &  0.00 &  0.00 \\
          &  0.37 &  1.05 \\
          &  1.13 &  2.66 \\
    \hline
    1.25  & -0.98 & -0.97 \\
          & -0.24 & -0.30 \\
          &  0.00 &  0.00 \\
          &  0.37 &  0.86 \\
          &  1.13 &  2.20 \\
          &  1.54 &  3.79 \\
    \hline
    1.50  & -0.24 & -0.17 \\
          &  0.00 &  0.00 \\
          &  0.37 &  0.82 \\
          &  1.13 &  1.68 \\
    \hline
    \end{tabular}
  \label{tab:addlabel}
\end{table}

\begin{figure}
\begin{center}
\includegraphics[scale=0.8, angle=0]{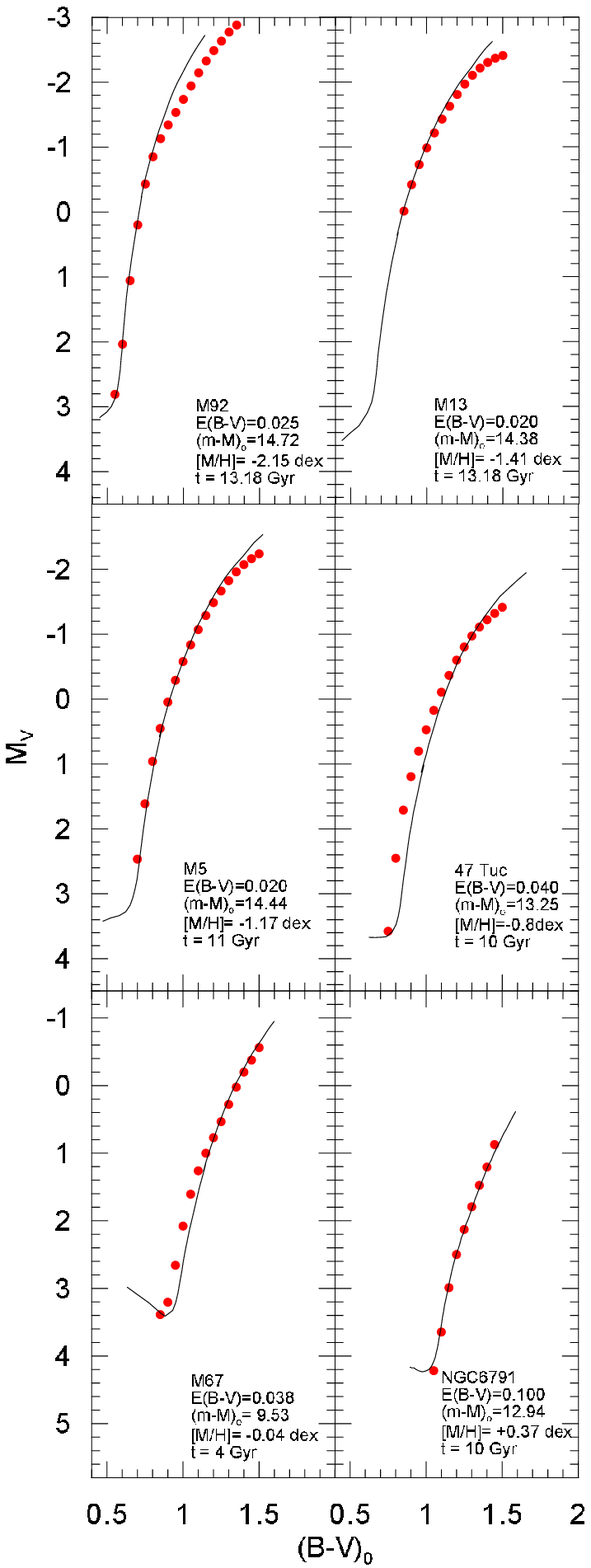} 
\caption[] {$M_{V}\times (B-V)_{0}$ absolute magnitude-colour diagrams fitted to isochrones.} 
\label{his:col}
\end{center}
\end {figure}

We adopted this interval in our study as $0.75\leq(B-V)_{0}\leq 1.50$ mag where at least three clusters are defined, and we evaluated $\Delta M$ offsets for each colour. Then, we combined them with the corresponding $\Delta[Fe/H]$ residuals and obtained the final calibrations. The general form of the equation for the calibrations is as follows:
\begin{eqnarray}
\Delta M=b_{0}+ b_{1}X + b_{2}X^{2}+ b_{3}X^{3}
\end{eqnarray}
where $X=\Delta[Fe/H]$.

$\Delta M$ could be fitted in terms of $\Delta[Fe/H]$ for the colours $0.75 \leq (B-V)_{0} \leq 0.82$ mag by a quadratic polynomial, whereas a cubic polynomial provided higher correlation for $(B-V)_{0}\geq $ 0.83 mag. The absolute magnitudes estimated via Eq. (2) for 76 colours and the corresponding $b_{i}$ ($i =$ 0,1,2,3) coefficients are given in Table 6. However, the numerical values for $\Delta M$ are omitted and the diagrams for the calibrations are not displayed for avoiding space consuming. One can use any data set taken from Table 6 depending on the required accuracy, and apply it to stars whose iron abundances are available.

The calibration of $\Delta M$ in terms of $\Delta[Fe/H]$ is carried out for the colour interval $0.75\leq (B-V)_{0} \leq1.50$ mag in steps of 0.1 mag. Small step is necessary to isolate an observational error on $B-V$ plus a wrong error due to reddening. The origin of the mentioned errors is the trend of the Red Giant Branch (RGB) sequence. As it is very steep, a small error in $B-V$ implies a large change in the absolute magnitude.

Iron abundance, $[Fe/H]$, is not the only parameter determining the chemistry of the star but also alpha enhancement, $[\alpha/Fe]$, is surely important. The $[\alpha/Fe]$ values for seven clusters used in our work are given in Table 1 (and Table 10, see Section 3.3). One can notice an inverse-correlation between two sets of abundances, i.e. $[\alpha/Fe]$ increases with decreasing $[Fe/H]$. Hence, we do not expect any considerable change in the numerical value of $\Delta M$ in the case of addition of an alpha enhancement term in Eq. (3).                 

\begin{table*}
\setlength{\tabcolsep}{2pt}
  \center
\tiny{
\caption{$M_{V}$ absolute magnitudes estimated for six Galactic clusters and the numerical values of $b_{i}$ ($i= 0, 1, 2, 3$) coefficients in Eq. (3). The last column gives the range of the metallicity $[Fe/H]$ (dex) for the star whose absolute magnitude would be estimated. R$^2$ is the square of the correlation coefficient.}
    \begin{tabular}{ccccccccccccc}
    \hline
          & M92   & M13   & M5    & 47 Tuc & M67   & NGC6791 &  &  &  &  &  &  \\
    \hline
    $(B-V)_{0}$ & $M_{V}$  & $M_{V}$  & $M_{V}$  & $M_{V}$  & $M_{V}$  & $M_{V}$  &$b_{0}$ & $b_{1}$ & $b_{2}$ & $b_{3}$ & R$^2$ & $[Fe/H]$-interval \\
    \hline
    0.75  & -0.43 &  $-$  & 1.61  & 3.58  &   $-$ & $-$   & 0.0000 & 4.4252 &  2.3862 &  $-$    & 1      & [-2,15, -0,80] \\
    0.76  & -0.53 &  $-$  & 1.47  & 3.31  &   $-$ & $-$   & 0.0000 & 4.1750 &  2.1787 &  $-$    & 1      & [-2,15, -0,80] \\
    0.77  & -0.62 &  $-$  & 1.33  & 3.07  &   $-$ & $-$   & 0.0000 & 3.9541 &  2.0013 &  $-$    & 1      & [-2,15, -0,80] \\
    0.78  & -0.71 &  $-$  & 1.20  & 2.84  &   $-$ & $-$   & 0.0000 & 3.7585 &  1.8506 &  $-$    & 1      & [-2,15, -0,80] \\
    0.79  & -0.78 &  $-$  & 1.08  & 2.64  &   $-$ & $-$   & 0.0000 & 3.5884 &  1.7260 &  $-$    & 1      & [-2,15, -0,80] \\
    0.80  & -0.85 &  $-$  & 0.96  & 2.45  &   $-$ & $-$   & 0.0000 & 3.4340 &  1.6174 &  $-$    & 1      & [-2,15, -0,80] \\
    0.81  & -0.92 &  $-$  & 0.85  & 2.28  &   $-$ & $-$   & 0.0000 & 3.2995 &  1.5280 &  $-$    & 1      & [-2,15, -0,80] \\
    0.82  & -0.98 &  $-$  & 0.74  & 2.12  &   $-$ & $-$   & 0.0000 & 3.1841 &  1.4602 &  $-$    & 1      & [-2,15, -0,80] \\
    0.83  & -1.03 &  0.19 & 0.64  & 1.97  &   $-$ & $-$   & 0.0000 & 2.3726 &  2.5943 &  1.9544 & 1      & [-2,15, -0,80] \\
    0.84  & -1.08 &  0.09 & 0.54  & 1.84  &   $-$ & $-$   & 0.0000 & 2.3826 &  2.3782 &  1.6756 & 1      & [-2,15, -0,80] \\
    0.85  & -1.13 & -0.01 & 0.45  & 1.71  & 3.39  & $-$   & 0.1142 & 2.8836 &  0.4740 & -0.7204 & 0.9982 & [-2,15, -0,04] \\
    0.86  & -1.18 & -0.11 & 0.36  & 1.59  & 3.40  & $-$   & 0.1012 & 2.8215 &  0.5387 & -0.6475 & 0.9987 & [-2,15, -0,04] \\
    0.87  & -1.22 & -0.19 & 0.28  & 1.48  & 3.39  & $-$   & 0.0912 & 2.7659 &  0.5891 & -0.5928 & 0.9989 & [-2,15, -0,04] \\
    0.88  & -1.26 & -0.27 & 0.20  & 1.38  & 3.35  & $-$   & 0.0837 & 2.7200 &  0.6255 & -0.5570 & 0.9991 & [-2,15, -0,04] \\
    0.89  & -1.30 & -0.35 & 0.12  & 1.29  & 3.28  & $-$   & 0.0777 & 2.6791 &  0.6510 & -0.5338 & 0.9992 & [-2,15, -0,04] \\
    0.90  & -1.34 & -0.42 & 0.05  & 1.19  & 3.20  & $-$   & 0.0738 & 2.6431 &  0.6668 & -0.5212 & 0.9993 & [-2,15, -0,04] \\
    0.91  & -1.38 & -0.49 & -0.03 & 1.11  & 3.11  & $-$   & 0.0714 & 2.6111 &  0.6750 & -0.5170 & 0.9993 & [-2,15, -0,04] \\
    0.92  & -1.42 & -0.55 & -0.10 & 1.03  & 3.00  & $-$   & 0.0700 & 2.5811 &  0.6764 & -0.5187 & 0.9993 & [-2,15, -0,04] \\
    0.93  & -1.46 & -0.61 & -0.16 & 0.95  & 2.89  & $-$   & 0.0694 & 2.5559 &  0.6727 & -0.5266 & 0.9993 & [-2,15, -0,04] \\
    0.94  & -1.50 & -0.67 & -0.23 & 0.87  & 2.78  & $-$   & 0.0700 & 2.5339 &  0.6654 & -0.5392 & 0.9993 & [-2,15, -0,04] \\
    0.95  & -1.54 & -0.73 & -0.29 & 0.80  & 2.66  & $-$   & 0.0711 & 2.5108 &  0.6542 & -0.5509 & 0.9992 & [-2,15, -0,04] \\
    0.96  & -1.58 & -0.78 & -0.35 & 0.73  & 2.54  & $-$   & 0.0721 & 2.4897 &  0.6408 & -0.5640 & 0.9992 & [-2,15, -0,04] \\
    0.97  & -1.62 & -0.84 & -0.41 & 0.66  & 2.42  & $-$   & 0.0730 & 2.4687 &  0.6264 & -0.5771 & 0.9991 & [-2,15, -0,04] \\
    0.98  & -1.66 & -0.89 & -0.47 & 0.60  & 2.30  & $-$   & 0.0750 & 2.4500 &  0.6108 & -0.5907 & 0.9990 & [-2,15, -0,04] \\
    0.99  & -1.70 & -0.94 & -0.52 & 0.54  & 2.19  & $-$   & 0.0763 & 2.4308 &  0.5947 & -0.6018 & 0.9990 & [-2,15, -0,04] \\
    1.00  & -1.74 & -0.99 & -0.58 & 0.47  & 2.08  & $-$   & 0.0776 & 2.4115 &  0.5787 & -0.6115 & 0.9989 & [-2,15, -0,04] \\
    1.01  & -1.78 & -1.03 & -0.63 & 0.41  & 1.97  & $-$   & 0.0792 & 2.3902 &  0.5631 & -0.6175 & 0.9988 & [-2,15, -0,04] \\
    1.02  & -1.82 & -1.08 & -0.69 & 0.35  & 1.87  & $-$   & 0.0797 & 2.3698 &  0.5483 & -0.6214 & 0.9987 & [-2,15, -0,04] \\
    1.03  & -1.86 & -1.13 & -0.74 & 0.29  & 1.78  & $-$   & 0.0811 & 2.3512 &  0.5351 & -0.6258 & 0.9987 & [-2,15, -0,04] \\
    1.04  & -1.90 & -1.17 & -0.79 & 0.23  & 1.69  & 4.25  & 0.0793 & 1.2145 &  0.4534 &  0.5178 & 0.9878 & [-2,15, +0,37] \\
    1.05  & -1.94 & -1.22 & -0.84 & 0.17  & 1.61  & 4.22  & 0.0792 & 1.1656 &  0.4407 &  0.5486 & 0.9872 & [-2,15, +0,37] \\
    1.06  & -1.99 & -1.26 & -0.89 & 0.12  & 1.53  & 4.14  & 0.0793 & 1.1393 &  0.4306 &  0.5586 & 0.9969 & [-2,15, +0,37] \\
    1.07  & -2.03 & -1.30 & -0.93 & 0.06  & 1.45  & 4.04  & 0.0797 & 1.1289 &  0.4226 &  0.5536 & 0.9869 & [-2,15, +0,37] \\
    1.08  & -2.07 & -1.35 & -0.98 & 0.00  & 1.38  & 3.92  & 0.0798 & 1.1318 &  0.4169 &  0.5363 & 0.9871 & [-2,15, +0,37] \\
    1.09  & -2.11 & -1.39 & -1.03 & -0.05 & 1.32  & 3.78  & 0.0790 & 1.1456 &  0.4138 &  0.5099 & 0.9875 & [-2,15, +0,37] \\
    1.10  & -2.15 & -1.43 & -1.07 & -0.11 & 1.26  & 3.64  & 0.0783 & 1.1630 &  0.4114 &  0.4802 & 0.9880 & [-2,15, +0,37] \\
    1.11  & -2.18 & -1.47 & -1.12 & -0.16 & 1.20  & 3.50  & 0.0780 & 1.1862 &  0.4111 &  0.4460 & 0.9886 & [-2,15, +0,37] \\
    1.12  & -2.22 & -1.51 & -1.16 & -0.21 & 1.15  & 3.37  & 0.0775 & 1.2102 &  0.4112 &  0.4119 & 0.9893 & [-2,15, +0,37] \\
    1.13  & -2.26 & -1.55 & -1.20 & -0.27 & 1.10  & 3.23  & 0.0764 & 1.2342 &  0.4128 &  0.3779 & 0.9899 & [-2,15, +0,37] \\
    1.14  & -2.29 & -1.59 & -1.25 & -0.32 & 1.05  & 3.11  & 0.0760 & 1.2573 &  0.4144 &  0.3455 & 0.9906 & [-2,15, +0,37] \\
    1.15  & -2.33 & -1.63 & -1.29 & -0.37 & 1.00  & 2.99  & 0.0756 & 1.2792 &  0.4168 &  0.3151 & 0.9912 & [-2,15, +0,37] \\
    1.16  & -2.36 & -1.67 & -1.33 & -0.42 & 0.95  & 2.88  & 0.0745 & 1.2978 &  0.4191 &  0.2876 & 0.9918 & [-2,15, +0,37] \\
    1.17  & -2.40 & -1.70 & -1.37 & -0.46 & 0.91  & 2.77  & 0.0736 & 1.3134 &  0.4211 &  0.2635 & 0.9923 & [-2,15, +0,37] \\
    1.18  & -2.43 & -1.74 & -1.41 & -0.51 & 0.86  & 2.67  & 0.0734 & 1.3258 &  0.4230 &  0.2422 & 0.9927 & [-2,15, +0,37] \\
    1.19  & -2.46 & -1.78 & -1.45 & -0.56 & 0.82  & 2.58  & 0.0730 & 1.3367 &  0.4252 &  0.2229 & 0.9931 & [-2,15, +0,37] \\
    1.20  & -2.49 & -1.81 & -1.49 & -0.60 & 0.77  & 2.50  & 0.0734 & 1.3423 &  0.4263 &  0.2078 & 0.9934 & [-2,15, +0,37] \\
    1.21  & -2.52 & -1.84 & -1.52 & -0.64 & 0.72  & 2.42  & 0.0731 & 1.3463 &  0.4263 &  0.1950 & 0.9937 & [-2,15, +0,37] \\
    1.22  & -2.55 & -1.88 & -1.56 & -0.69 & 0.68  & 2.34  & 0.0735 & 1.3483 &  0.4256 &  0.1844 & 0.9939 & [-2,15, +0,37] \\
    1.23  & -2.58 & -1.91 & -1.60 & -0.73 & 0.63  & 2.27  & 0.0739 & 1.3469 &  0.4236 &  0.1761 & 0.9940 & [-2,15, +0,37] \\
    1.24  & -2.61 & -1.94 & -1.63 & -0.77 & 0.58  & 2.20  & 0.0747 & 1.3444 &  0.4205 &  0.1695 & 0.9941 & [-2,15, +0,37] \\
    1.25  & -2.64 & -1.97 & -1.67 & -0.80 & 0.53  & 2.13  & 0.0755 & 1.3394 &  0.4165 &  0.1646 & 0.9941 & [-2,15, +0,37] \\
    1.26  & -2.66 & -2.00 & -1.70 & -0.84 & 0.48  & 2.06  & 0.0770 & 1.3328 &  0.4114 &  0.1611 & 0.9941 & [-2,15, +0,37] \\
    1.27  & -2.69 & -2.03 & -1.73 & -0.88 & 0.43  & 1.99  & 0.0784 & 1.3274 &  0.4053 &  0.1575 & 0.9940 & [-2,15, +0,37] \\
    1.28  & -2.72 & -2.05 & -1.76 & -0.91 & 0.38  & 1.92  & 0.0796 & 1.3179 &  0.3975 &  0.1564 & 0.9939 & [-2,15, +0,37] \\
    1.29  & -2.75 & -2.08 & -1.80 & -0.94 & 0.33  & 1.86  & 0.0818 & 1.3102 &  0.3889 &  0.1548 & 0.9937 & [-2,15, +0,37] \\
    1.30  & -2.77 & -2.11 & -1.83 & -0.97 & 0.28  & 1.79  & 0.0840 & 1.3004 &  0.3794 &  0.1545 & 0.9935 & [-2,15, +0,37] \\
    1.31  & -2.80 & -2.13 & -1.85 & -1.00 & 0.23  & 1.73  & 0.0862 & 1.2898 &  0.3701 &  0.1542 & 0.9933 & [-2,15, +0,37] \\
    1.32  & -2.82 & -2.15 & -1.88 & -1.03 & 0.17  & 1.66  & 0.0881 & 1.2792 &  0.3604 &  0.1542 & 0.9930 & [-2,15, +0,37] \\
    1.33  & -2.85 & -2.18 & -1.91 & -1.06 & 0.12  & 1.60  & 0.0906 & 1.2670 &  0.3507 &  0.1546 & 0.9927 & [-2,15, +0,37] \\
    1.34  & -2.87 & -2.20 & -1.94 & -1.09 & 0.07  & 1.53  & 0.0934 & 1.2547 &  0.3425 &  0.1546 & 0.9923 & [-2,15, +0,37] \\
    1.35  & -2.88 & -2.22 & -1.96 & -1.11 & 0.02  & 1.47  & 0.0957 & 1.2417 &  0.3355 &  0.1542 & 0.9919 & [-2,15, +0,37] \\
    1.36  & -2.89 & -2.24 & -1.99 & -1.14 & -0.03 & 1.42  & 0.0975 & 1.2280 &  0.3312 &  0.1527 & 0.9915 & [-2,15, +0,37] \\
    1.37  & -2.90 & -2.25 & -2.01 & -1.16 & -0.07 & 1.36  & 0.1002 & 1.2118 &  0.3297 &  0.1510 & 0.9910 & [-2,15, +0,37] \\
    1.38  & -2.90 & -2.27 & -2.03 & -1.18 & -0.12 & 1.31  & 0.1018 & 1.1942 &  0.3321 &  0.1473 & 0.9905 & [-2,15, +0,37] \\
    1.39  &   $-$ & -2.29 & -2.05 & -1.20 & -0.16 & 1.26  & 0.1766 & 1.7354 & -1.0652 &  0.8117 & 0.9931 & [-1,41, +0,37] \\
    1.40  &   $-$ & -2.30 & -2.07 & -1.22 & -0.20 & 1.20  & 0.1794 & 1.7251 & -1.0742 &  0.8125 & 0.9927 & [-1,41, +0,37] \\
    1.41  &   $-$ & -2.32 & -2.09 & -1.24 & -0.24 & 1.15  & 0.1814 & 1.7150 & -1.0787 &  0.8103 & 0.9924 & [-1,41, +0,37] \\
    1.42  &   $-$ & -2.33 & -2.11 & -1.26 & -0.28 & 1.10  & 0.1829 & 1.7047 & -1.0739 &  0.8010 & 0.9921 & [-1,41, +0,37] \\
    1.43  &   $-$ & -2.35 & -2.13 & -1.28 & -0.31 & 1.03  & 0.1837 & 1.6942 & -1.0561 &  0.7812 & 0.9918 & [-1,41, +0,37] \\
    1.44  &   $-$ & -2.36 & -2.15 & -1.30 & -0.35 & 0.96  & 0.1833 & 1.6817 & -1.0186 &  0.7470 & 0.9915 & [-1,41, +0,37] \\
    1.45  &   $-$ & -2.37 & -2.16 & -1.32 & -0.38 & $-$   & 0.0000 & 1.6233 &  2.6509 & -2.3794 & 1      & [-1,41, -0,04] \\
    1.46  &   $-$ & -2.38 & -2.18 & -1.34 & -0.41 & $-$   & 0.0000 & 1.6073 &  2.6821 & -2.4063 & 1      & [-1,41, -0,04] \\
    1.47  &   $-$ & -2.39 & -2.20 & -1.36 & -0.45 & $-$   & 0.0000 & 1.5920 &  2.7009 & -2.4241 & 1      & [-1,41, -0,04] \\
    1.48  &   $-$ & -2.40 & -2.21 & -1.38 & -0.48 & $-$   & 0.0000 & 1.5742 &  2.7399 & -2.4585 & 1      & [-1,41, -0,04] \\
    1.49  &   $-$ & -2.40 & -2.23 & -1.40 & -0.52 & $-$   & 0.0000 & 1.5530 &  2.7816 & -2.4954 & 1      & [-1,41, -0,04] \\
    1.50  &   $-$ & -2.41 & -2.24 & -1.42 & -0.56 & $-$   & 0.0000 & 1.5327 &   2.8113 & -2.527 & 1      & [-1,41, -0,04] \\

\hline    
\end{tabular}
}
  \label{tab:addlabel}
\end{table*}
\subsection{Absolute Magnitude as a Function of Metallicity and Age}

Age plays an important role in the trend of the fiducial sequence of the RGB. Hence, we added age as a parameter in the calibration of the absolute magnitude as follows:

\begin{eqnarray}
   M_{V} = c_{0} + c_{1}x + c_{2}y + c_{3}x^{2} + c_{4}y^{2} + c_{5}xy
\end{eqnarray}
where $x$ and $y$ indicate the metallicity and age, i.e. $x=[Fe/H]$ and $y=t$. The metallicities and ages of the clusters can be used and the coefficients $c_{i}$ ($i=$ 0, 1, 2, 3, 4, 5) can be determined. The common domain of the clusters is $1.05\leq (B-V)_{0}\leq 1.35$ mag. Hence, this procedure works only for the $(B-V)_{0}$ colour interval just cited. However, we could extend this procedure to the interval 0.85 $\leq (B-V)_{0} <$ 1.05 mag. by a small modification of Eq. (4) as in the following:

\begin{eqnarray}
   M_{V} = d_{0} + d_{1}x + d_{2}y + d_{3}x^{2} + d_{4}xy
\end{eqnarray}
Thus, we omitted the fifth term in Eq. (4) and obtained Eq. (5) with five coefficients, i.e. $d_{i}$ ($i=$ 0, 1, 2, 3, 4), which can be determined by using the metallicities and ages of the clusters M92, M13, M5, 47 Tuc, and M67. The elimination of the fifth term is due to fact that its coefficient ($c_{4}$) is the (absolutely) smallest one (see Section 3.3). One can use Eq. (4) and Eq. (5) simultaneously and estimate the absolute magnitudes of red giants with 0.85 $\leq (B-V)_{0} <$ 1.35. 

We used the metallicities of the clusters M92, M13, M5, 47 Tuc, M67, and NGC 6791 in Table 1, their ages in Table 7 and the absolute magnitudes of these clusters in Table 4 for $(B-V)_{0}=$ 1.05, 1.10, 1.15, 1.20, 1.25, 1.30, 1.35, and determined  seven sets of coefficients $c_{i}$ ($i =$ 0, 1, 2, 3, 4, 5). The coefficients $d_{i}$ ($i =$ 0, 1, 2, 3, 4) were determined by the same procedure for the same $(B-V)_{0}$ colour-indices omitting the data of NGC 6791, however. The results are given in Table 8 and Table 9. The aim of the limitation for the $(B-V)_{0}$ colour is to test the effect of age on absolute magnitude estimation. One can extend the determination of the coefficients $c_{i}$ and $d_{i}$ to a larger $(B-V)_{0}$ interval, accordingly.     

\subsection{Application of the Method}
\subsubsection{Absolute Magnitudes Estimated by means of Metallicity}

We applied the method to seven clusters with different metallicities, i.e. M3, M53, M71, NGC 188, NGC 6366, IC 4499, and Ter 7, as explained in the following. The reason of choosing clusters instead of individual field giants is that clusters provide absolute magnitudes for comparison with the ones estimated by means of our method. The distance modulus, colour excess and metallicity of the clusters are given in Table 10, whereas the $V$ and $B-V$ data are presented in Table 11 and they are calibrated in Fig. 5. The data in Table 10 and Table 11 are taken from the authors cited in the reference list of Table 10. In the case of two references, the first one refers to the $V$, $(B-V)$ fiducial sequence, whereas the second one refers to the data given in the same table.

\begin{table}
  \centering
\scriptsize{
  \caption{Ages of the clusters. The ages of the clusters used for the absolute magnitude calibration, M92, M13, M5, 47 Tuc, M67 and NGC 6791 are determined in this study as showed Fig. 3. Whereas,  the ages of the clusters used for the application of the procedure are taken from the authors given in the reference list.}
    \begin{tabular}{lc|lcc}
    \hline
    Cluster & t (Gyr) & Cluster & t (Gyr) & Ref. \\
    \hline
    M92     & 13.18 & M3       & 12.1  & 1 \\
    M13     & 13.18 & M53      & 13.2  & 2 \\
    M5      & 11.00 & M71      & 10.1  & 1 \\
    47 Tuc  & 10.00 & NGC 188  & 5.9   & 3 \\
    M67     & 4.00  & NGC 6366 & 8.0   & 1 \\
    NGC 6791& 10.00 & IC 4499  & 11.2  & 1 \\
    $-$     & $-$   & Ter 7    & 8.9   & 1 \\
    \hline
    \end{tabular}\\
(1) \cite{Salaris02}, (2) \cite{Santos04}, (3) \cite{Meibom09}
}
  \label{tab:addlabel}
\end{table}

\begin{figure}
\begin{center}
\includegraphics[scale=0.6, angle=0]{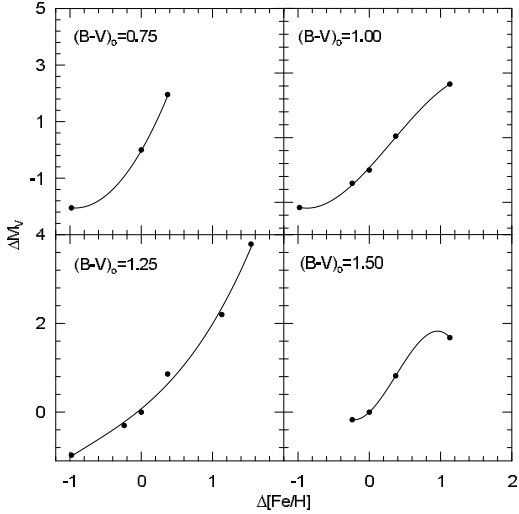} 
\caption[] {Calibration of the absolute magnitude offsets ($\Delta M$) as a function of metallicity offsets ($\Delta [Fe/H]$) for four colour-indices.} 
\label{his:col}
\end{center}
\end {figure}

\begin{table}
  \centering
  \caption{Numerical values of the coefficients in Eq. (4) for seven $(B-V)_{0}$ colour indices.}
    \begin{tabular}{ccccccc}
    \hline
 $(B-V)_{0}$ & $c_{0}$ & $c_{1}$& $c_{2}$    & $c_{3}$    & $c_{4}$    & $c_{5}$ \\
    \hline
      1.05  & -3.62 & 12.00 & 1.88  & -0.08 & -0.12 & -0.86 \\
      1.10  & -3.73 & 11.48 & 1.80  & -0.13 & -0.12 & -0.83 \\
      1.15  & -3.95 & 11.54 & 1.80  & -0.28 & -0.12 & -0.88 \\
      1.20  & -4.14 & 11.58 & 1.79  & -0.37 & -0.12 & -0.91 \\
      1.25  & -4.51 & 12.19 & 1.84  & -0.54 & -0.13 & -0.99 \\
      1.30  & -5.04 & 12.26 & 1.94  & -0.56 & -0.13 & -1.01 \\
      1.35  & -5.7  & 12.84 & 2.09  & -0.69 & -0.14 & -1.09 \\
    \hline
    \end{tabular}
  \label{tab:addlabel}
\end{table}

\begin{table}
  \centering
  \caption{Numerical values of the coefficients in Eq. (5) for seven $(B-V)_{0}$ colour indices.}
    \begin{tabular}{ccccccc}
    \hline
    $(B-V)_{0}$ & $d_{0}$ & $d_{1}$& $d_{2}$ & $d_{3}$ & $d_{4}$ \\
    \hline
    1.05  &  0.13 & 4.18  & 0.42  & 1.82  & 0.25 \\
    1.10  & -0.13 & 3.96  & 0.39  & 1.69  & 0.23 \\
    1.15  & -0.27 & 3.86  & 0.36  & 1.59  & 0.21 \\
    1.20  & -0.42 & 3.80  & 0.34  & 1.52  & 0.19 \\
    1.25  & -0.65 & 3.77  & 0.34  & 1.49  & 0.18 \\
    1.30  & -0.96 & 3.75  & 0.35  & 1.51  & 0.19 \\
    1.35  & -1.32 & 3.69  & 0.38  & 1.53  & 0.20 \\
    \hline
    \end{tabular}
  \label{tab:addlabel}
\end{table}

As in the case of Table 2, the $(V, B-V)$ points in Table 11 consist of the fiducial sequence of the referred cluster. Hence, they represent the cluster in question quite well. However, they are not error free. Although one can expect a bit larger error for the photographic data of the cluster M3, this did not show up (see Table 12, column 7). 

We evaluated the $\Delta M$ absolute magnitude offsets by using the Eq. (3) for the $(B-V)_{0}$-domain of each cluster, added them to the corresponding absolute magnitude of the cluster M5 and obtained the absolute magnitude $(M_{V})_{*}$. The results are presented in Table 12. The columns give: (1) $(B-V)_{0}$ colour index, (2) $M_V$, absolute magnitude for the cluster estimated by its colour magnitude diagram, (3) $(M_{V})_{M5}$, absolute magnitude corresponding to the cluster M5, (4) $\Delta[Fe/H]$, metallicity offset from the metallicity of the cluster M5, (5) $\Delta M$, absolute magnitude offset from the fiducial giant sequence of the cluster M5, (6) $(M_{V})_{*}$, the absolute magnitude estimated by the procedure, (7) absolute magnitude residuals: $M_V-(M_{V})_{*}$. The cluster name is indicated at the top of the corresponding columns.   

The differences between the absolute magnitudes estimated in this study and the ones evaluated via the colour magnitude diagrams of the clusters, i.e. $M_V-(M_{V})_{*}$, lie between $-0.61$ and $+0.66$ mag. However, the range of 91\% of the residuals is shorter, i.e. $-0.4 \leq M_{V}-(M_{V})_{*} \leq + 0.4$ mag, and their mean and standard deviation are only 0.05 and 0.19 mag, respectively. The positive large values correspond to the data of cluster NGC 188, whereas the negative ones originate from the data of cluster NGC 6366. The results are given in Table 13 and Fig. 6. 
\begin{figure}[h]
\begin{center}
\includegraphics[scale=0.5, angle=0]{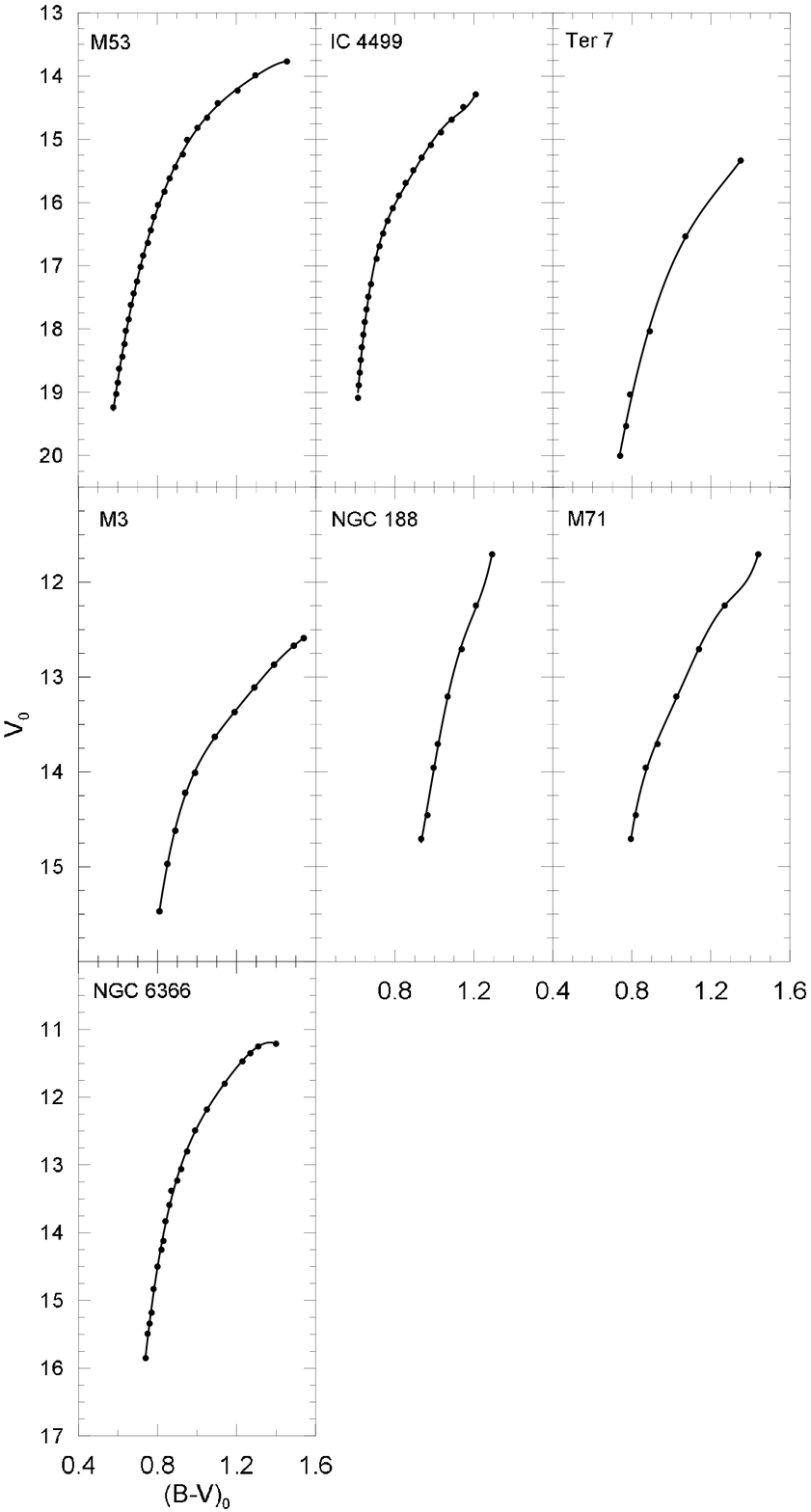}
\caption[] {$V_{0}\times(B-V)_{0}$ colour-apparent magnitude diagrams for the Galactic clusters used for the application of the procedure.} 
\end{center}
\end{figure}  

\begin{figure}[h]
\begin{center}
\includegraphics[angle=0, width=80mm, height=50mm]{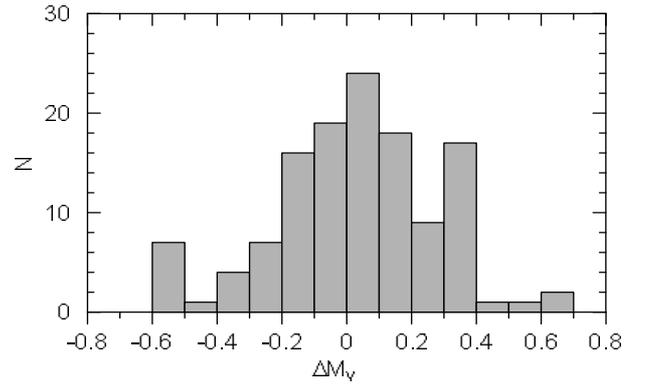}
\caption{Histogram of the residuals.}
\label{histogram}
\end{center}
\end{figure}  

\begin{table}
\setlength{\tabcolsep}{2pt}
\begin{center}
\scriptsize{
  \caption{Data for the clusters used for the application of the method.}
    \begin{tabular}{lcccccl}
    \hline
    Cluster & $V-M_{V}$ & $E(B-V)$ & $(V-M_{V})_{0}$ & $[Fe/H]$ & $[\alpha/Fe]$ &Reference \\
            & (mag) & (mag) & (mag) & (dex) & (dex) &  \\
    \hline
    M3      & 15.07 & 0.010  & 15.04 & -1.50 & 0.29 &(1),(2)\\
    M53     & 16.32 & 0.020  & 16.26 & -1.99 & $-$  &(3),(2)\\
    M71     & 13.70 & 0.280  & 12.83 & -0.78 & 0.39 &(4)    \\
    NGC188  & 11.40 & 0.095  & 11.11 & -0.01 & $-$  &(5)    \\
    NGC6366 &  $-$  & 0.700  & 12.26 & -0.67 & $-$  &(6)    \\
    IC4499  & 17.08 & 0.230  & 16.37 & -1.53 & $-$  &(7),(2)\\
    Ter7    & 17.01 & 0.070  & 16.79 & -0.87 & 0.009 &(8)    \\
    \hline
    \end{tabular}\\
}
(1) \cite{Sandage70}, (2) \citet{Harris96,Harris10}, (3) \cite{Rey98}, (4) \cite{Hodder92}, (5) \cite{Meibom09}, (6) \cite{Alonso97}, (7) \cite{Sarajedini93}, (8) \cite{Buonanno95}\\
\end{center}
  \end{table}

\begin{table*}
  \centering
  \caption{Fiducial giant sequences for the Galactic clusters used in the application of the procedure.}
    \begin{tabular}{cccc|cccc|cccc}
    \hline
    $B-V$ & $V$ & $(B-V)_{0}$ & $V_{0}$ & $B-V$ & $V$ & $(B-V)_{0}$ & $V_{0}$ & $B-V$ & $V$ & $(B-V)_{0}$ & $V_{0}$\\
    \hline
    \multicolumn{4}{c|}{M3}       & \multicolumn{4}{c|}{M71}      & \multicolumn{4}{c}{IC 4499} \\
\hline
		1.550 & 12.62 & 1.540 & 12.59 & 1.720 & 12.50 & 1.440 & 11.63 & 1.440 & 15.00 & 1.210 & 14.29 \\
		1.500 & 12.70 & 1.490 & 12.67 & 1.550 & 13.00 & 1.270 & 12.13 & 1.377 & 15.20 & 1.147 & 14.49 \\
		1.400 & 12.90 & 1.390 & 12.87 & 1.420 & 13.50 & 1.140 & 12.63 & 1.318 & 15.40 & 1.088 & 14.69 \\
		1.300 & 13.14 & 1.290 & 13.11 & 1.305 & 14.00 & 1.025 & 13.13 & 1.264 & 15.60 & 1.034 & 14.89 \\
		1.200 & 13.40 & 1.190 & 13.37 & 1.210 & 14.50 & 0.930 & 13.63 & 1.213 & 15.80 & 0.983 & 15.09 \\
		1.100 & 13.66 & 1.090 & 13.63 & 1.150 & 15.00 & 0.870 & 14.13 & 1.166 & 16.00 & 0.936 & 15.29 \\
		1.000 & 14.04 & 0.990 & 14.01 & 1.100 & 15.50 & 0.820 & 14.63 & 1.124 & 16.20 & 0.894 & 15.49 \\
		0.950 & 14.25 & 0.940 & 14.22 & 1.075 & 16.00 & 0.795 & 15.13 & 1.085 & 16.40 & 0.855 & 15.69 \\
		0.900 & 14.65 & 0.890 & 14.62 & 1.060 & 16.50 & 0.780 & 15.63 & 1.051 & 16.60 & 0.821 & 15.89 \\
		0.860 & 15.00 & 0.850 & 14.97 & 1.040 & 17.00 & 0.760 & 16.13 & 1.020 & 16.80 & 0.790 & 16.09 \\
\cline{5-8}     0.820 & 15.50 & 0.810 & 15.47 & \multicolumn{4}{c|}{NGC 188}  & 0.994 & 17.00 & 0.764 & 16.29 \\
\cline{1-8}    \multicolumn{4}{c|}{M53}       & 1.387 & 12.00 & 1.292 & 11.71 & 0.971 & 17.20 & 0.741 & 16.49 \\
\cline{1-4}     1.476 & 13.83 & 1.456 & 13.77 & 1.305 & 12.54 & 1.210 & 12.25 & 0.953 & 17.40 & 0.723 & 16.69 \\
		1.316 & 14.05 & 1.296 & 13.99 & 1.233 & 13.00 & 1.138 & 12.71 & 0.938 & 17.60 & 0.708 & 16.89 \\
		1.226 & 14.29 & 1.206 & 14.23 & 1.162 & 13.50 & 1.067 & 13.21 & 0.910 & 18.00 & 0.680 & 17.29 \\
		1.126 & 14.49 & 1.106 & 14.43 & 1.112 & 14.00 & 1.017 & 13.71 & 0.896 & 18.20 & 0.666 & 17.49 \\
		1.072 & 14.72 & 1.052 & 14.66 & 1.091 & 14.25 & 0.996 & 13.96 & 0.887 & 18.40 & 0.657 & 17.69 \\
		1.023 & 14.88 & 1.003 & 14.82 & 1.060 & 14.75 & 0.965 & 14.46 & 0.878 & 18.60 & 0.648 & 17.89 \\
		0.971 & 15.07 & 0.951 & 15.01 & 1.028 & 15.00 & 0.933 & 14.71 & 0.871 & 18.80 & 0.641 & 18.09 \\
\cline{5-8}     0.949 & 15.30 & 0.929 & 15.24 & \multicolumn{4}{c|}{NGC 6366} & 0.863 & 19.00 & 0.633 & 18.29 \\
\cline{5-8}     0.911 & 15.50 & 0.891 & 15.44 & 2.100 & 13.38 & 1.400 & 11.21 & 0.858 & 19.20 & 0.628 & 18.49 \\
		0.882 & 15.68 & 0.862 & 15.62 & 2.010 & 13.42 & 1.310 & 11.25 & 0.853 & 19.40 & 0.623 & 18.69 \\
		0.856 & 15.89 & 0.836 & 15.83 & 1.970 & 13.52 & 1.270 & 11.35 & 0.848 & 19.60 & 0.618 & 18.89 \\
		0.823 & 16.10 & 0.803 & 16.04 & 1.930 & 13.64 & 1.230 & 11.47 & 0.844 & 19.80 & 0.614 & 19.09 \\
\cline{9-12}    0.802 & 16.29 & 0.782 & 16.23 & 1.840 & 13.97 & 1.140 & 11.80 & \multicolumn{4}{c}{Terzan7}   \\
\cline{9-12}    0.787 & 16.50 & 0.767 & 16.44 & 1.750 & 14.35 & 1.050 & 12.18 & 1.770 & 15.05 & 1.700 & 14.83 \\
		0.772 & 16.70 & 0.752 & 16.64 & 1.690 & 14.66 & 0.990 & 12.49 & 1.420 & 15.55 & 1.350 & 15.33 \\
		0.748 & 16.90 & 0.728 & 16.84 & 1.650 & 14.97 & 0.950 & 12.80 & 1.140 & 16.75 & 1.070 & 16.53 \\
		0.735 & 17.08 & 0.715 & 17.02 & 1.620 & 15.23 & 0.920 & 13.06 & 0.960 & 18.25 & 0.890 & 18.03 \\
		0.717 & 17.31 & 0.697 & 17.25 & 1.600 & 15.40 & 0.900 & 13.23 & 0.860 & 19.25 & 0.790 & 19.03 \\
		0.700 & 17.50 & 0.680 & 17.44 & 1.570 & 15.55 & 0.870 & 13.38 & 0.840 & 19.75 & 0.770 & 19.53 \\
		0.686 & 17.68 & 0.666 & 17.62 & 1.560 & 15.76 & 0.860 & 13.59 & 0.810 & 20.22 & 0.740 & 20.00 \\
		0.675 & 17.91 & 0.655 & 17.85 & 1.540 & 16.00 & 0.840 & 13.83 & $-$   & $-$   & $-$   & $-$   \\
		0.660 & 18.09 & 0.640 & 18.03 & 1.530 & 16.29 & 0.830 & 14.12 & $-$   & $-$   & $-$   & $-$   \\
		0.654 & 18.30 & 0.634 & 18.24 & 1.520 & 16.42 & 0.820 & 14.25 & $-$   & $-$   & $-$   & $-$   \\
		0.643 & 18.50 & 0.623 & 18.44 & 1.500 & 16.67 & 0.800 & 14.50 & $-$   & $-$   & $-$   & $-$   \\
		0.626 & 18.69 & 0.606 & 18.63 & 1.480 & 17.00 & 0.780 & 14.83 & $-$   & $-$   & $-$   & $-$   \\
		0.620 & 18.91 & 0.600 & 18.85 & 1.470 & 17.35 & 0.770 & 15.18 & $-$   & $-$   & $-$   & $-$   \\
		0.612 & 19.09 & 0.592 & 19.03 & 1.460 & 17.51 & 0.760 & 15.34 & $-$   & $-$   & $-$   & $-$   \\
		0.597 & 19.30 & 0.577 & 19.24 & 1.450 & 17.66 & 0.750 & 15.49 & $-$   & $-$   & $-$   & $-$   \\
		$-$   & $-$   & $-$   & $-$   & 1.440 & 18.02 & 0.740 & 15.85 & $-$   & $-$   & $-$   & $-$   \\
    \hline
     \end{tabular}
  \label{tab:addlabel}
\end{table*}

\begin{table*}
\setlength{\tabcolsep}{2pt}
\center
\tiny{
\caption{$(M_{V})_{*}$ absolute magnitudes and the residuals estimated by the procedure explained in our work. The columns give: (1) $(B-V)_{0}$ colour index, (2) $M_V$, absolute magnitude for the cluster estimated by its colour magnitude diagram, (3) $(M_{V})_{M5}$, absolute magnitude corresponding to the cluster M5, (4) $\Delta[Fe/H]$, metallicity offset from the metallicity of cluster M5, (5) $\Delta M$, absolute magnitude offset from the fiducial giant sequence of cluster M5, (6) $(M_{V})_{*}$, the absolute magnitude estimated by the procedure, (7) absolute magnitude residuals: $M_V-(M_{V})_{*}$.}
    \begin{tabular}{ccccccc|ccccccc}
    \hline
    1     & 2     & 3     & 4     & 5     & 6     & 7     & 1     & 2     & 3     & 4     & 5     & 6     & 7 \\
    \hline
    $(B-V)_{0}$ &$M_{V}$&$(M_{V})_{M5}$& $\Delta[Fe/H]$ & $\Delta M$ & $(M_{V})_{*}$ & $(2)-(6)$ & $(B-V)_{0}$ &$M_{V}$&$(M_{V})_{M5}$& $\Delta[Fe/H]$ & $\Delta M$ & $(M_{V})_{*}$ & $(2)-(6)$\\
    \hline
    \multicolumn{7}{c|}{M3}            & \multicolumn{7}{c}{M53 (cont.)}                     \\
    \hline
    0.85  & -0.07 & 0.45  & -0.33 & -0.76 & -0.31 & 0.24  & 0.97  & -1.28 & -0.41 & -0.82 & -1.21 & -1.62 & 0.34 \\
    0.90  & -0.52 & 0.05  & -0.33 & -0.71 & -0.66 & 0.14  & 1.00  & -1.42 & -0.58 & -0.82 & -1.17 & -1.75 & 0.33 \\
    0.95  & -0.85 & -0.29 & -0.33 & -0.67 & -0.96 & 0.10  & 1.02  & -1.50 & -0.69 & -0.82 & -1.15 & -1.84 & 0.34 \\
    1.00  & -1.10 & -0.58 & -0.33 & -0.63 & -1.21 & 0.11  & 1.05  & -1.61 & -0.84 & -0.82 & -0.88 & -1.72 & 0.10 \\
    1.05  & -1.28 & -0.84 & -0.33 & -0.28 & -1.11 & -0.17 & 1.10  & -1.78 & -1.07 & -0.82 & -0.86 & -1.93 & 0.16 \\
    1.10  & -1.44 & -1.07 & -0.33 & -0.28 & -1.35 & -0.09 & 1.12  & -1.84 & -1.16 & -0.82 & -0.87 & -2.03 & 0.19 \\
    1.15  & -1.58 & -1.29 & -0.33 & -0.31 & -1.60 & 0.02  & 1.15  & -1.92 & -1.29 & -0.82 & -0.87 & -2.15 & 0.23 \\
    1.20  & -1.71 & -1.49 & -0.33 & -0.33 & -1.82 & 0.11  & 1.17  & -1.97 & -1.37 & -0.82 & -0.87 & -2.23 & 0.26 \\
    1.25  & -1.84 & -1.67 & -0.33 & -0.33 & -1.99 & 0.16  & 1.20  & -2.04 & -1.49 & -0.82 & -0.86 & -2.34 & 0.30 \\
    1.30  & -1.96 & -1.83 & -0.33 & -0.31 & -2.13 & 0.17  & 1.22  & -2.09 & -1.56 & -0.82 & -0.85 & -2.41 & 0.32 \\
    1.35  & -2.09 & -1.96 & -0.33 & -0.28 & -2.24 & 0.16  & 1.25  & -2.16 & -1.67 & -0.82 & -0.83 & -2.50 & 0.34 \\
    1.40  & -2.20 & -2.07 & -0.33 & -0.54 & -2.61 & 0.40  & 1.27  & -2.20 & -1.73 & -0.82 & -0.82 & -2.56 & 0.36 \\
    1.45  & -2.31 & -2.16 & -0.33 & -0.16 & -2.33 & 0.02  & 1.30  & -2.26 & -1.83 & -0.82 & -0.81 & -2.64 & 0.38 \\
    1.50  & -2.40 & -2.24 & -0.33 & -0.11 & -2.35 & -0.05 & 1.32  & -2.30 & -1.88 & -0.82 & -0.80 & -2.69 & 0.39 \\
\cline{1-7}    \multicolumn{7}{c|}{M71}                   & 1.35  & -2.35 & -1.96 & -0.82 & -0.78 & -2.74 & 0.39 \\
\cline{1-7}0.80  & 2.20  & 0.96  & 0.39  & 1.59  & 2.54  & -0.34 & 1.37  & -2.39 & -2.01 & -0.82 & -0.76 & -2.76 & 0.38\\
\cline{8-14}0.85  & 1.40  & 0.45  & 0.39  & 1.27  & 1.72  & -0.32 & \multicolumn{7}{c}{Ter7}                       \\
\cline{8-14}    0.90  & 0.93  & 0.05  & 0.39  & 1.18  & 1.22  & -0.29 & 0.75  & 3.01  & 1.61  & 0.30  & 1.54  & 3.15  & -0.15 \\
    0.95  & 0.63  & -0.29 & 0.39  & 1.12  & 0.83  & -0.19 & 0.77  & 2.69  & 1.33  & 0.30  & 1.37  & 2.70  & -0.01 \\
    1.00  & 0.39  & -0.58 & 0.39  & 1.07  & 0.49  & -0.10 & 0.80  & 2.25  & 0.96  & 0.30  & 1.18  & 2.13  & 0.11 \\
    1.05  & 0.15  & -0.84 & 0.39  & 0.63  & -0.20 & 0.35  & 0.82  & 1.98  & 0.74  & 0.30  & 1.09  & 1.83  & 0.15 \\
    1.10  & -0.10 & -1.07 & 0.39  & 0.62  & -0.45 & 0.34  & 0.85  & 1.60  & 0.45  & 0.30  & 1.00  & 1.45  & 0.15 \\
    1.15  & -0.35 & -1.29 & 0.39  & 0.66  & -0.63 & 0.28  & 0.90  & 1.06  & 0.05  & 0.30  & 0.91  & 0.96  & 0.11 \\
    1.20  & -0.57 & -1.49 & 0.39  & 0.67  & -0.81 & 0.24  & 0.92  & 0.87  & -0.10 & 0.30  & 0.89  & 0.80  & 0.08 \\
    1.25  & -0.74 & -1.67 & 0.39  & 0.67  & -1.00 & 0.26  & 0.95  & 0.61  & -0.29 & 0.30  & 0.87  & 0.58  & 0.03 \\
    1.30  & -0.85 & -1.83 & 0.39  & 0.66  & -1.17 & 0.31  & 0.97  & 0.45  & -0.41 & 0.30  & 0.85  & 0.44  & 0.00 \\
    1.35  & -0.94 & -1.96 & 0.39  & 0.64  & -1.32 & 0.38  & 1.00  & 0.22  & -0.58 & 0.30  & 0.84  & 0.26  & -0.04 \\
    1.40  & -1.08 & -2.07 & 0.39  & 0.74  & -1.34 & 0.26  & 1.02  & 0.08  & -0.69 & 0.30  & 0.82  & 0.14  & -0.06 \\
    1.45  & -1.41 & -2.16 & 0.39  & 0.90  & -1.27 & -0.14 & 1.05  & -0.11 & -0.84 & 0.30  & 0.48  & -0.35 & 0.24  \\
\cline{1-7}    \multicolumn{7}{c|}{NGC188}                & 1.10  & -0.40 & -1.07 & 0.30  & 0.48  & -0.59 & 0.19 \\
\cline{1-7}1.00  & 2.79  & -0.58 & 1.17  & 2.71  & 2.13  & 0.66  & 1.12  & -0.51 & -1.16 & 0.30  & 0.49  & -0.67 & 0.16 \\
    1.02  & 2.56  & -0.69 & 1.17  & 2.61  & 1.92  & 0.63  & 1.15  & -0.66 & -1.29 & 0.30  & 0.51  & -0.78 & 0.12 \\
    1.05  & 2.23  & -0.84 & 1.17  & 2.92  & 2.09  & 0.14  & 1.17  & -0.75 & -1.37 & 0.30  & 0.51  & -0.86 & 0.10 \\
    1.07  & 2.03  & -0.93 & 1.17  & 2.87  & 1.93  & 0.10  & 1.20  & -0.89 & -1.49 & 0.30  & 0.52  & -0.97 & 0.08 \\
    1.10  & 1.77  & -1.07 & 1.17  & 2.77  & 1.70  & 0.07  & 1.22  & -0.97 & -1.56 & 0.30  & 0.52  & -1.04 & 0.07 \\
    1.12  & 1.62  & -1.16 & 1.17  & 2.72  & 1.56  & 0.07  & 1.25  & -1.10 & -1.67 & 0.30  & 0.52  & -1.15 & 0.05 \\
    1.15  & 1.43  & -1.29 & 1.17  & 2.65  & 1.36  & 0.07  & 1.27  & -1.18 & -1.73 & 0.30  & 0.52  & -1.22 & 0.04 \\
    1.17  & 1.31  & -1.37 & 1.17  & 2.61  & 1.24  & 0.07  & 1.30  & -1.29 & -1.83 & 0.30  & 0.51  & -1.31 & 0.02 \\
    1.20  & 1.16  & -1.49 & 1.17  & 2.56  & 1.07  & 0.08  & 1.32  & -1.36 & -1.88 & 0.30  & 0.51  & -1.37 & 0.01 \\
    1.22  & 1.05  & -1.56 & 1.17  & 2.53  & 0.97  & 0.08  & 1.35  & -1.47 & -1.96 & 0.30  & 0.50  & -1.46 & -0.01 \\
    1.25  & 0.87  & -1.67 & 1.17  & 2.48  & 0.81  & 0.06  & 1.37  & -1.53 & -2.01 & 0.30  & 0.50  & -1.51 & -0.02 \\
    1.27  & 0.73  & -1.73 & 1.17  & 2.44  & 0.71  & 0.02  & 1.40  & -1.63 & -2.07 & 0.30  & 0.62  & -1.45 & -0.17 \\
    1.30  & 0.44  & -1.83 & 1.17  & 2.37  & 0.55  & -0.11 & 1.42  & -1.68 & -2.11 & 0.30  & 0.62  & -1.49 & -0.19 \\
\cline{1-7}    \multicolumn{7}{c|}{IC4499}                & 1.45  & -1.77 & -2.16 & 0.30  & 0.66  & -1.50 & -0.26 \\
\cline{1-14}0.80  & -0.34 & 0.96  & -0.36 & -1.03 & -0.07 & -0.28 & \multicolumn{7}{c}{NGC 6366}                          \\
\cline{8-14}0.82  & -0.46 & 0.74  & -0.36 & -0.96 & -0.22 & -0.24 & 0.85  & 1.50  & 0.45  & 0.50  & 1.58  & 2.04 &-0.54 \\
    0.85  & -0.62 & 0.45  & -0.36 & -0.83 & -0.38 & -0.25 & 0.90  & 0.94  & 0.05  & 0.50  & 1.50  & 1.54  & -0.61 \\
    0.90  & -0.88 & 0.05  & -0.36 & -0.77 & -0.72 & -0.16 & 0.92  & 0.76  & -0.10 & 0.50  & 1.46  & 1.37  & -0.61 \\
    0.92  & -0.99 & -0.10 & -0.36 & -0.75 & -0.84 & -0.14 & 0.95  & 0.52  & -0.29 & 0.50  & 1.42  & 1.13  & -0.61 \\
    0.95  & -1.14 & -0.29 & -0.36 & -0.72 & -1.01 & -0.13 & 0.97  & 0.38  & -0.41 & 0.50  & 1.39  & 0.98  & -0.60 \\
    0.97  & -1.24 & -0.41 & -0.36 & -0.71 & -1.12 & -0.12 & 1.00  & 0.20  & -0.58 & 0.50  & 1.35  & 0.77  & -0.57 \\
    1.00  & -1.37 & -0.58 & -0.36 & -0.69 & -1.27 & -0.10 & 1.02  & 0.09  & -0.69 & 0.50  & 1.32  & 0.64  & -0.55 \\
    1.02  & -1.45 & -0.69 & -0.36 & -0.67 & -1.36 & -0.09 & 1.05  & -0.06 & -0.84 & 0.50  & 0.84  & 0.00  & -0.07 \\
    1.05  & -1.56 & -0.84 & -0.36 & -0.31 & -1.14 & -0.41 & 1.10  & -0.29 & -1.07 & 0.50  & 0.82  & -0.25 & -0.04 \\
    1.07  & -1.62 & -0.93 & -0.36 & -0.30 & -1.23 & -0.39 & 1.12  & -0.37 & -1.16 & 0.50  & 0.84  & -0.32 & -0.05 \\
    1.10  & -1.69 & -1.07 & -0.36 & -0.31 & -1.38 & -0.31 & 1.15  & -0.50 & -1.29 & 0.50  & 0.86  & -0.43 & -0.07 \\
    1.12  & -1.74 & -1.16 & -0.36 & -0.32 & -1.48 & -0.26 & 1.17  & -0.57 & -1.37 & 0.50  & 0.87  & -0.50 & -0.07 \\
    1.15  & -1.81 & -1.29 & -0.36 & -0.35 & -1.63 & -0.18 & 1.20  & -0.68 & -1.49 & 0.50  & 0.88  & -0.61 & -0.07 \\
    1.17  & -1.87 & -1.37 & -0.36 & -0.36 & -1.73 & -0.14 & 1.22  & -0.75 & -1.56 & 0.50  & 0.88  & -0.68 & -0.07 \\
    1.20  & -2.00 & -1.49 & -0.36 & -0.36 & -1.85 & -0.14 & 1.25  & -0.85 & -1.67 & 0.50  & 0.87  & -0.80 & -0.05 \\
    1.22  & -2.12 & -1.56 & -0.36 & -0.37 & -1.93 & -0.20 & 1.27  & -0.91 & -1.73 & 0.50  & 0.86  & -0.87 & -0.04 \\
\cline{1-7}    \multicolumn{7}{c|}{M53}                   & 1.30  & -0.98 & -1.83 & 0.50  & 0.85  & -0.98 & 0.00  \\
\cline{1-7}0.85  & -0.56 & 0.45  & -0.82 & -1.53 & -1.08 & 0.52  & 1.32  & -1.02 & -1.88 & 0.50  & 0.84  & -1.04 & 0.03\\
    0.90  & -0.90 & 0.05  & -0.82 & -1.36 & -1.31 & 0.41  & 1.35  & -1.05 & -1.96 & 0.50  & 0.82  & -1.14 & 0.09 \\
    0.92  & -1.02 & -0.10 & -0.82 & -1.31 & -1.40 & 0.38  & 1.37  & -1.06 & -2.01 & 0.50  & 0.81  & -1.20 & 0.14 \\
    0.95  & -1.18 & -0.29 & -0.82 & -1.24 & -1.53 & 0.35  & 1.40  & -1.05 & -2.07 & 0.50  & 0.87  & -1.20 & 0.15 \\
    \hline
\end{tabular}
}
 \label{tab:addlabel}
\end{table*}
\subsubsection{Absolute Magnitudes Estimated by Means of Metallicity and Age}

We applied the procedure presented in Section 3.2 to the same clusters, i.e. M3, M53, M71, NGC 188, NGC 6366, IC 4499, and Ter 7, and estimated two sets of absolute magnitudes for the colour indices $(B-V)_{0}=$ 1.05, 1.10, 1.15, 1.20, 1.25, 1.30, 1.35. We used the coefficients $c_{i}$ and $d_{i}$ in Table 8 and Table 9 for the two sets of absolute magnitudes. The ages and metallicities of the clusters used in the corresponding equations are taken from Table 7 and Table 10. The results are given in Table 14 and Table 15.

\begin{table}[h]
  \centering
  \caption{Distribution of residuals. $N$ denotes the number of stars.}
    \begin{tabular}{rcc}
    \hline
    $\Delta M$-interval & $<\Delta M>$ & N \\
    \hline
    (-0.6,-0.5] & -0.58 & 7  \\
    (-0.5,-0.4] & -0.41 & 1  \\
    (-0.4,-0.3] & -0.34 & 4  \\
    (-0.3,-0.2] & -0.25 & 7  \\
    (-0.2,-0.1] & -0.15 & 16 \\
     (-0.1,0.0] & -0.05 & 19 \\
     (0.0, 0.1] &  0.06 & 24 \\
     (0.1, 0.2] &  0.15 & 18 \\
     (0.2, 0.3] &  0.26 & 9  \\
     (0.3, 0.4] &  0.36 & 17 \\
     (0.4, 0.5] &  0.41 & 1  \\
     (0.5, 0.6] &  0.52 & 1  \\
     (0.6, 0.7] &  0.65 & 2  \\
    \hline
    \end{tabular}
  \label{residuals}
\end{table}

\begin{table}
  \centering
\scriptsize{
 \caption{Absolute magnitudes (second column) and residuals (fourth column) estimated by the procedure explained in Section 3.2. The coefficients $c_{i}$ ($i=$0, 1, 2, 3, 4, 5) are taken from Table 8. The absolute magnitude $M_{cl}$ (taken from Table 12) indicates the absolute magnitude estimated by the colour magnitude diagram of the corresponding cluster.}

    \begin{tabular}{clccc}
    \hline
$(B-V)_{0}$ & Cluster & $M_{ev}$ & $M_{cl}$ & $\Delta M$\\
    \hline
    1.05 & M3      & -1.38 & -1.28 & 0.10  \\
    1.05 & M53     & -1.77 & -1.61 & 0.16  \\
    1.05 & M71     & 0.25  & 0.15  & -0.10 \\
    1.05 & NGC 188 & 2.53  & 2.23  & -0.30 \\
    1.05 & NGC 6366& 0.28  & -0.06 & -0.34 \\
    1.05 & IC 4499 & -1.42 & -1.56 & -0.14 \\
    1.05 & Ter 7   & -0.23 & -0.11 & 0.12  \\
    \hline
    1.10 & M3      & -1.61 & -1.44 & 0.17  \\
    1.10 & M53     & -2.00 & -1.78 & 0.22  \\
    1.10 & M71     & -0.05 & -0.10 & -0.05 \\
    1.10 & NGC 188 & 2.13  & 1.77  & -0.36 \\
    1.10 & NGC 6366& -0.16 & -0.29 & -0.13 \\
    1.10 & IC 4499 & -1.96 & -1.69 & 0.27  \\
    1.10 & Ter 7   & -0.69 & -0.40 & 0.29  \\
    \hline
    1.15 & M3      & -1.71 & -1.58 & 0.13  \\
    1.15 & M53     & -2.06 & -1.92 & 0.14  \\
    1.15 & M71     & -0.25 & -0.35 & -0.10 \\
    1.15 & NGC 188 & 1.86  & 1.43  & -0.43 \\
    1.15 & NGC 6366& -0.37 & -0.50 & -0.13 \\
    1.15 & IC 4499 & -2.07 & -1.81 & 0.26  \\
    1.15 & Ter 7   & -0.87 & -0.66 & 0.21  \\
    \hline
    1.20 & M3      & -1.74 & -1.71 & 0.03  \\
    1.20 & M53     & -2.03 & -2.04 & -0.01 \\
    1.20 & M71     & -0.39 & -0.57 & -0.18 \\
    1.20 & NGC 188 & 1.62  & 1.16  & -0.46 \\
    1.20 & NGC 6366& -0.55 & -0.68 & -0.13 \\
    1.20 & IC 4499 & -2.13 & -2.00 & 0.13  \\
    1.20 & Ter 7   & -1.24 & -0.89 & 0.35  \\
    \hline
    1.25 & M3      & -2.81 & -1.84 & 0.97  \\
    1.25 & M53     & -3.26 & -2.16 & 1.10  \\
    1.25 & M71     & -1.22 & -0.74 & 0.48  \\
    1.25 & NGC 188 & 1.18  &  0.87 & -0.31 \\
    1.25 & NGC 6366& -1.21 & -0.85 & 0.36  \\
    1.25 & Ter 7   & -2.02 & -1.10 & 0.92  \\
    \hline
    1.30 & M3      & -1.92 & -1.96 & -0.04 \\
    1.30 & M53     & -2.17 & -2.26 & -0.09 \\
    1.30 & M71     & -0.65 & -0.85 & -0.20 \\
    1.30 & NGC 188 & 1.25  &  0.44 & -0.80 \\
    1.30 & NGC 6366& -0.89 & -0.98 & -0.09 \\
    1.30 & Ter 7   & -1.59 & -1.29 &  0.30 \\
    \hline
    1.35 & M3      & -1.94 & -2.09 & -0.15 \\
    1.35 & M53     & -2.16 & -2.35 & -0.19 \\
    1.35 & M71     & -0.72 & -0.94 & -0.22 \\
    1.35 & NGC 6366& -1.01 & -1.05 & -0.04 \\
    1.35 & Ter 7   & -1.71 & -1.47 &  0.24 \\
 \hline
    \end{tabular}
}
  \label{tab:addlabel}
\end{table}

\begin{table}
  \centering
\scriptsize{
 \caption{Absolute magnitudes (second column) and residuals (fourth column) estimated by the procedure explained in Section 3.2. The coefficients $d_{i}$ ($i=$ 0, 1, 2, 3, 4) are taken from Table 9. The absolute magnitude $M_{cl}$ (taken from Table 12) indicates the absolute magnitude estimated by the colour magnitude diagram of the corresponding cluster.}
    \begin{tabular}{clccc}
    \hline
$(B-V)_{0}$ & Cluster & $M_{ev}$ & $M_{cl}$ & $\Delta M$\\
  \hline
    1.05 & M3       & -1.47 & -1.28 & 0.19 \\
    1.05 & M53      & -1.96 & -1.61 & 0.35 \\
    1.05 & M71      & 0.27  & 0.15  & -0.12\\
    1.05 & NGC 6366 & 0.18  & -0.06 & -0.24\\
    1.05 & IC 4499  & -1.55 & -1.56 & -0.01\\
    1.05 & Ter 7    & -0.31 & -0.11 & 0.20 \\
    1.05 & NGC 188  & 2.07  & 2.23  & 0.16 \\
    \hline
    1.10 & M3       & -1.67 & -1.44 & 0.23 \\
    1.10 & M53      & -2.15 & -1.78 & 0.37 \\
    1.10 & M71      & -0.02 & -0.10 & -0.08\\
    1.10 & NGC 6366 & -0.10 & -0.29 & -0.19\\
    1.10 & IC 4499  & -1.76 & -1.69 & 0.07 \\
    1.10 & Ter 7    & -0.56 & -0.40 & 0.16 \\
    1.10 & NGC 188  & 1.69  & 1.77  & 0.08 \\
    \hline
    1.15 & M3       & -1.86 & -1.58 & 0.28 \\
    1.15 & M53      & -2.33 & -1.92 & 0.41 \\
    1.15 & M71      & -0.28 & -0.35 & -0.07\\
    1.15 & NGC 6366 & -0.35 & -0.50 & -0.15\\
    1.15 & IC 4499  & -1.95 & -1.81 & 0.14 \\
    1.15 & Ter 7    & -0.80 & -0.66 & 0.14 \\
    1.15 & NGC 188  & 1.38  & 1.43  & 0.05 \\
    \hline
    1.20 & M3       & -2.04 & -1.71 & 0.33 \\
    1.20 & M53      & -2.48 & -2.04 & 0.44 \\
    1.20 & M71      & -0.51 & -0.57 & -0.06\\
    1.20 & NGC 6366 & -0.57 & -0.68 & -0.11\\
    1.20 & IC 4499  & -2.13 & -2.00 & 0.13 \\
    1.20 & Ter 7    & -1.02 & -0.89 & 0.13 \\
    1.20 & NGC 188  & 1.12  & 1.16  & 0.04 \\
    \hline
    1.25 & M3       & -2.20 & -1.84 & 0.36 \\
    1.25 & M53      & -2.63 & -2.16 & 0.47 \\
    1.25 & M71      & -0.72 & -0.74 & -0.02\\
    1.25 & NGC 6366 & -0.78 & -0.85 & -0.07\\
    1.25 & Ter 7    & -1.21 & -1.10 & 0.11 \\
    1.25 & NGC 188  & 0.88  & 0.87  & -0.01\\
    \hline
    1.30 & M3       & -2.35 & -1.96 & 0.39 \\
    1.30 & M53      & -2.77 & -2.26 & 0.51 \\
    1.30 & M71      & -0.88 & -0.85 & 0.03 \\
    1.30 & NGC 6366 & -0.98 & -0.98 & 0.00 \\
    1.30 & Ter 7    & -1.40 & -1.29 & 0.11 \\
    1.30 & NGC 188  & 0.66  & 0.44  & -0.22\\
    \hline
    1.35 & M3       & -2.47 & -2.09 & 0.38 \\
    1.35 & M53      & -2.88 & -2.35 & 0.53 \\
    1.35 & M71      & -1.02 & -0.94 & 0.08 \\
    1.35 & NGC 6366 & -1.15 & -1.05 & 0.10 \\
    1.35 & Ter 7    & -1.55 & -1.47 & 0.08 \\
    \hline
    \end{tabular}
}
  \label{tab:addlabel}
\end{table}

We compared the absolute magnitude residuals evaluated via three procedures for the colour indices $(B-V)_{0}=$ 1.05, 1.10, 1.15, 1.20, 1.25, 1.30, 1.35 which are given in Tables 12, 14 and 15, in order to choose the most advantage one and to test the effect of the age in absolute magnitude estimation of the red giants. The best statistic is the couple of their mean and standard deviation for this purpose. Table 16 shows that the three procedures provide absolute magnitudes agreeable with the ones appeared in the literature. However, the range of the residuals for the absolute magnitudes evaluated by the procedures which  involve age as a term are larger. That is, the procedure where the absolute magnitude is fitted to a third degree polynomial of the metallicity has an advantage respect to the others two. Hence, the procedures with large range of absolute magnitude residuals have not been extended to the other $(B-V)_{0}$ colour indices.   

The absolute magnitudes on the RGB at a given colour and metallicity do not change linearly or quadratically with age, as implied by the form of Eqs. (4) and (5). Instead, the absolute magnitude gets rapidly fainter for young (and so massive) stars with a certain $B-V$ and $[Fe/H]$, but shows virtually the same absolute magnitude for all old stars, i.e. $t>6$ Gyr. 

\begin{table}
\centering
\caption{Mean ($<\Delta M>$) and standard deviation ($\sigma$) for the absolute magnitude residuals for three procedures. (1), (2), and (3) in the first column refer to the procedures given in Sections 3.3.1, 3.3.2 with coefficients $c_{i}$, and Section 3.3.2 but with coefficients $d_{i}$} 
    \begin{tabular}{ccc}
    \hline
    Procedure & $<\Delta M>$  & $\sigma$\\
    \hline
    1     & 0.12  & 0.20 \\
    2     & 0.04  & 0.36 \\
    3     & 0.13  & 0.21 \\
    \hline
    \end{tabular}
  \label{tab:addlabel}
\end{table}

\section{Summary and Discussion}
We presented an absolute magnitude calibration for red giants based on the colour--magnitude diagrams of six Galactic clusters with different metallicities, i.e. M92, M13, M5, 47 Tuc, M67, and NGC 6791. We combined the calibrations between $V_{0}$ and $(B-V)_{0}$ for each cluster with their true distance modulus and evaluated a set o absolute magnitudes for the $(B-V)_{0}$ range of each clusters. We adopted the $V_{0}, (B-V)_{0}$ calibration of the cluster M5 which is defined in the interval $0.75 \leq (B-V)_{0} \leq 1.50$ mag as the standard colour magnitude diagram and evaluated the $\Delta M$ offsets from the fiducial red giant sequence of the cluster M5. We, then combined $\Delta M$ with the corresponding $\Delta[Fe/H]$ offsets and obtained the required calibration.

We applied the procedure to another set of Galactic cluster, i.e. M3, M53, M71, NGC 188, NGC 6366, IC 4499, and Ter 7. The reason of this choice is that a cluster provides absolute magnitude for comparison with the ones estimated by means of our procedure. We used the calibration in Eq. (3) and evaluated a set of $\Delta M$ offsets for each cluster in their $(B-V)_{0}$ range. We, then added them to the corresponding absolute magnitudes derived for the standard cluster M5 and obtained the required absolute magnitudes. We compared the absolute magnitudes estimated by this procedure with those evaluated via combination of the fiducial $V_{0}$, $(B-V)_{0}$ sequence and the true distance modulus for each cluster. 91\% of the differences between two sets of absolute magnitudes (the residuals) lie in the range (-0.4, +0.4). The mean and the standard deviation of the residuals are 0.05 and 0.19 mag. The residuals cited in this study are at the level of the ones appeared in the literature. We quote two works as example to confirm our argument, i.e. \cite{Laird88} and \cite{Karaali03}.

Comparison of our work with the one of \cite{Hog98} shows that there is an improvement on our results with respect to theirs. \cite{Hog98} fitted the absolute magnitudes of 581 bright K giants in terms of two colours defined in the DDO intermediate band photometry \citep{McClure76}, i.e. C4245 and C4548, by a quadratic polynomial. Then, they applied two small corrections to obtain more accurate absolute magnitudes. The range of the residuals of the final absolute magnitudes with respect to the {\em Hipparcos} absolute magnitudes is [-1, +1]. They give an accuracy of 0.35 mag which confirm the advantage of our results. We quote also the work of \cite{Ljunggren66}.

Although age plays an important role in the trend of the fiducial sequence of the RGB, we have not used it as a parameter in this calibration of the absolute magnitude. Another problem may originate from the Red Clump (RC) stars. These stars lie very close to the RGB but they present a completely different group of stars. Table 13 and Fig. 6 summarize how reliable are our absolute magnitudes. If age and possibly the mix with RC stars would affect our results this should show up. Additionally, we should add that the fiducial sequences used in our study were properly selected as RGB.  However, the researches should identify and exclude the RC stars when they apply our calibrations to the field stars. 

Despite the considerations stated in the preceding paragraph, we fitted the absolute magnitude to metallicity and age for a limited sub-sample of $(B-V)_{0}$ colour, i.e. 1.05, 1.10, 1.15, 1.20, 1.25, 1.30, 1.35, and compared the residuals evaluated by this procedure with the corresponding ones evaluated by means of the former procedure. The results in Table 16 confirm our argument. That is, the procedure where the absolute magnitude fitted to a third degree polynomial of (only) the metallicity provides more reliable absolute magnitudes than the procedure which involves the age as a parameter.      
   
The colour magnitude diagrams displayed in Fig. 2 are smooth and confirm the dependence of the absolute magnitude calibration in Eq. (3) on colour and metallicity. The absolute magnitudes of a cluster with lower martallicty are brighter for given a colour. This argument was used as a tool to prefer the iron abundance $[Fe/H]=$ -1.17 to -1.40 dex for the cluster M5, where these values were proposed by \cite{Sandquist96}, and \cite{Zinn84}, respectively. The colour magnitude diagram of M5 is about 0.5 mag fainter than the absolute magnitude of M13, for a given colour. Hence, the metallicities of these clusters should be different. As the iron abundance of M13 appeared in the literature is $[Fe/H]=-1.41$ dex \citep{Gratton97}, the one of M5 should be different than this value.

One requires accurate metallicity and interstellar extinction determination for the application of the procedure to the field stars. The absolute magnitude could be calibrated as a function of ultraviolet excess, instead of metallicity. However, accurate ultraviolet magnitudes can not be provided easily. Whereas, metallicity can be derived by different methods, such as by means of atmospheric parameters of a star, a procedure which is applied rather extensively in large surveys such as RAdial Velocity Experiment \citep[RAVE;][]{Steinmetz06}. In such cases, one needs to transform the calibration from {\em BV} to the system in question. The clusters considered in our paper are relatively old, $t\geq4$ Gyr. For such stars the age is a secondary parameter and does not influence much the position of the RGB sequence. However, the field stars can be much younger. We should remind that the derived relations are applicable only to stars older than 4 Gyr.

\section*{Acknowledgments}
We thank to the anonymous reviewer for his/her comments and suggestions. This research has made use of NASA's Astrophysics Data System and the SIMBAD database, operated at CDS, Strasbourg, France.

\end{document}